\documentclass[12pt]{article}

\usepackage{epsfig}
\usepackage{cite}
\usepackage{amsmath, amssymb, amsfonts}
\usepackage{color}
\usepackage{latexsym}  
\usepackage{graphicx}
\usepackage{cancel}
\usepackage[colorlinks,bookmarks]{hyperref}
\hypersetup{pdfpagemode=UseNone, pdfstartview=FitH, linkcolor=blue, citecolor=red, urlcolor=blue}

\def\be{\begin{equation}}
\def\ee{\end{equation}}
\def\ba{\begin{eqnarray}}
\def\ea{\end{eqnarray}}

\def\bdm{\begin{displaymath}}
\def\edm{\end{displaymath}}
\def\la{~\mbox{\raisebox{-.6ex}{$\stackrel{<}{\sim}$}}~}
\def\ga{~\mbox{\raisebox{-.6ex}{$\stackrel{>}{\sim}$}}~}
\def\bq{\begin{quote}}
\def\eq{\end{quote}}

 at 10truept

\newcommand{\p}{\partial}

\renewcommand{\[}{\left[}
\renewcommand{\]}{\right]}

\newcommand{\Mpl}{M_{\mathrm{Pl}}}

\newcommand{\bea}{\begin{eqnarray}}
\newcommand{\eea}{\end{eqnarray}}

\newcommand{\bi}{\begin{itemize}}
\newcommand{\ei}{\end{itemize}}

\newcommand{\beq}{\begin{equation}}
\newcommand{\eeq}{\end{equation}}
\newcommand{\beqa}{\begin{eqnarray}}
\newcommand{\eeqa}{\end{eqnarray}}
\newcommand{\mpl}{\Mpl}

\def\la{~\mbox{\raisebox{-.6ex}{$\stackrel{<}{\sim}$}}~}
\def\ga{~\mbox{\raisebox{-.6ex}{$\stackrel{>}{\sim}$}}~}

\def\12{{1 \over 2}}

\def\ltap{\ \raise.3ex\hbox{$<$\kern-.75em\lower1ex\hbox{$\sim$}}\ }
\def\gtap{\ \raise.3ex\hbox{$>$\kern-.75em\lower1ex\hbox{$\sim$}}\ }
\def\gl{\ \raise.5ex\hbox{$>$}\kern-.8em\lower.5ex\hbox{$<$}\ }
\def\roughly#1{\raise.3ex\hbox{$#1$\kern-.75em\lower1ex\hbox{$\sim$}}}

\begin{document}

\thispagestyle{empty}
\begin{flushright}
May 2025 
\end{flushright}
\vspace*{1.35cm}
\begin{center}

 {\Large \bf A Quantal Theory of Restoration of} 
 \vskip.3cm
 {\Large \bf Strong CP Symmetry}  

\vspace*{1cm} {\large 
Nemanja Kaloper\footnote{\tt
kaloper@physics.ucdavis.edu}
}\\
\vspace{.5cm}
{\em QMAP, Department of Physics and Astronomy, University of
California}\\
\vspace{.05cm}
{\em Davis, CA 95616, USA}\\

\vspace{1.5cm} ABSTRACT
\end{center}
We propose a mechanism for relaxing 
a gauge theory CP violating phase in discrete steps to very small values. The
idea is that the CP violating phase includes the magnetic dual of a $4$-form flux which 
can discharge by the nucleation of membranes. 
Inside the bubbles surrounded by the membranes, 
the total CP violating phase is reduced. When the bubbles are produced rapidly during
radiation domination in the early universe, near the chiral symmetry breaking scale, 
they will collide and percolate, melting away into gauge theory 
radiation and dramatically relaxing CP violation.

\vfill \setcounter{page}{0} \setcounter{footnote}{0}

\vspace{1cm}
\newpage

\section{Introduction}

What if there is no QCD axion? Or, what if some UV physics disrupts the axion too much 
to solve the strong CP problem of QCD? Is a smoothly varying axion field even necessary to 
relax the strong CP-violating phase?
Here we put forward an alternative to axion for dynamically relaxing the strong CP violation, 
based on discrete reduction of the CP-breaking phase $\theta_{\cancel{\tt CP}}$ 
in the early universe, which minimizes the $\theta_{\cancel{\tt CP}}$-dependent 
contribution to the total energy of the QCD vacuum below chiral symmetry breaking scale.
This mechanism of cosmological relaxation by the production of percolating bubbles of the 
CP-conserving phase around the time of chiral symmetry breaking may be an example of how
$\theta_{\cancel{\tt CP}}$ could be relaxed cosmologically \cite{Weinberg:1978uk}.

Our proposal is inspired by an insight provided in the paper \cite{Aurilia:1980xj}. In 
1980, Aurilia, Nicolai and Townsend noted that both the strong CP violating phase 
$\theta_{\cancel{\tt CP}}$ and the cosmological constant\footnote{The 
same conclusion about the cosmological constant was also noted separately in \cite{Aurilia:1978qs,Duff:1980qv}.} 
are degenerate with a magnetic dual of a nonvanishing $4$-form flux. This follows from the dynamics 
of a  strongly coupled $SU(N)$ gauge theory in the large $N$-limit, below chiral symmetry breaking, 
where it can be described by a nonlinear $\sigma$-model with CP violation, with new
extended degrees of freedom 
\cite{Coleman:1976uz,Rosenzweig:1979ay,DiVecchia:1980yfw,Nath:1979ik,Witten:1980sp,Ohta:1981ai}. 
These degrees of freedom are membranes charged under a $4$-form field strength (a.k.a. a ``top'' 
form, the term designating that its rank is the same as the number of dimensions).  
L\"uscher teased the top form Lagrangian out from the nonperturbative contributions 
to the correlation functions of the duals of gauge theory anomaly terms \cite{Luscher:1978rn} even earlier.
These results were studied and confirmed in two-dimensional $CP^{n-1}$ $\sigma$-models at 
low energies in more detail \cite{DAdda:1978vbw,Witten:1978bc}.

Including explicitly the $4$-form sector of the theory in the action provides a very direct
and lucid tool for understanding both the CP violation and the possible solutions for restoring CP.
Dvali has advocated this approach forcefully in several papers 
\cite{Dvali:2003br,Dvali:2004tma,Dvali:2005an,Dvali:2005zk,Dvali:2007iv}, explaining 
that the axion solution of the CP problem corresponds to Higgsing
the top form gauge theory \cite{Dvali:2005an}. A similar viewpoint was also presented long ago
by Aurilia, Takahashi and Townsend \cite{Aurilia:1980jz}. Moreover, Dvali has also noted 
that a membrane discharge dynamics could address the problem even when the axion is 
absent \cite{Dvali:2005zk}. While it has been noted that non-supersymmetric large-N gauge theories 
have a landscape of infinitely many non-degenerate vacua \cite{Jackiw:1976pf,Callan:1976je}, 
concerns were raised that the decay rates are slow, both in the large N limit \cite{Shifman:1998if} and
for smaller gauge group ranks like in QCD \cite{Forbes:2000et}. 
To enhance the membrane nucleation rates, \cite{Dvali:2005zk} proposed that the 
membrane charge depends on the top form flux, and that
$\theta_{\cancel{\tt CP}} = 0$ is a very strong attractor 
for the terminal flux. Further, explicit 
examples \cite{Dvali:1998ms,Dubovsky:2011tu} exist where the 
barrier to tunneling diminishes as N grows large. 

Here we take a more mundane approach which allows us to 
circumvent this subtle issue altogether, as outlined in \cite{Kaloper:2025wgn}. 
We will show that when the strong CP-violating phase is degenerate with a magnetic dual of the flux of a new
top form, which is a replica of L\"uscher's top form that emerges below chiral symmetry breaking in QCD, 
the discharges which relax $\theta_{\cancel{\tt CP}}$ occur. Adopting a bottom-up approach, 
we treat the charge and tension of the membranes that source the new top form as input 
parameters, whose properties are not directly determined by QCD dynamics. We select their
values so that the nucleations of membranes allow the
fluxes in question to relax fast. Since the new top form flux is degenerate with
the flux of L\"uscher's top form, the discharge cascade simultaneously reduces both, and so it 
relaxes the strong CP violation. This occurs near the QCD chiral symmetry breaking scale,
when the universe is radiation dominated, expanding slowly. Therefore the bubbles of true vacuum
inside the membranes will expand at the speed of light, colliding with each other.
Upon collisions, the membranes separating them will decay into the strongly-coupled 
QCD excitations. This conflagration of membranes yields the percolation of the 
true vacuum regions, which have a tiny $\theta_{\cancel{\tt CP}}$, meaning that the
CP violation would be reduced over a huge swath of the universe. 

The mechanism is completely analogous to the discharge of a Maxwell $U(1)$ field by 
pair production. We will work out the discharge rates in the decoupling limit of gravity 
when $\mpl \rightarrow \infty$, treating the background spacetime as locally Minkowski when membranes 
nucleate, which works since the dynamics of nucleations is covariant, and the membrane radius at nucleation 
is very small. This will suffice for our purposes. Hence we will ignore any implications or interplay of the strong 
CP problem and the cosmological constant conundrum, fine tuning the latter at will to be as small as
required. 

Since the relaxation of  $\theta_{\cancel{\tt CP}}$  toward small values occurs in discrete steps, controlled
by the membrane charges, generically the typical terminal value of the CP-violating phase $\theta_{\cancel{\tt CP}}$ 
will not be exactly zero. Therefore, the neutron dipole moment could naturally be non-vanishing. Further, the 
relaxation dynamics is similar to first order phase transitions near the QCD scale, and so bubble collisions might 
also produce a signal in the stochastic gravity wave background \cite{Kosowsky:1992rz,Kamionkowski:1993fg}. Since
the nucleations are very prolific, determining precise abundance of those gravity waves requires a more 
careful study. There could be additional effects, which we outline in the discussion below. 

\section{Emergent Top Form Sector} 

Nontrivial topology of the gauge sector in QCD forces an extension of the perturbative Lagrangian,
which must include an extra topological term 
\be
{\cal L}_{\tt QCD} \ni \frac{g^2}{64\pi^2} \theta \epsilon^{\mu\nu\lambda\sigma}  
\sum_a G^a_{\mu\nu} G^a_{\lambda\sigma} \, , 
\label{cpterm}
\ee
with a canonically normalized gauge field strength 
$G^a$. Here $\sum_a$ is the sum 
over colors, and $\theta$ is a constant ``vacuum angle", $\theta \in [0,2\pi]$. Note, that aside from the 
$1/g^2$ factor in the gauge kinetic term, we follow the normalizations used in \cite{Luscher:1978rn}.

This term is the chiral anomaly, and is required to ensure that the theory is 
invariant under large gauge transformations. It 
arises even in the pure glue limit, when the quarks are absent. This term leads to the resolution 
of the $U(1)$ problem and the explanation of the 
missing Goldstone meson \cite{Veneziano:1979ec,Kogut:1974kt}. 
The $\theta$ parameter is \`a priori arbitrary; if there were no 
other contributions to (\ref{cpterm}), a nonzero $\theta$ 
would yield CP violation. However, in the regime where the theory has chiral symmetry, unbroken 
in the fermion sector, all the values of $\theta$ 
are degenerate, implying that the theory has many 
different, topologically distinct superselection sectors.

One way to see this is to note that this term is purely topological, 
being a total derivative. Indeed, using the gauge potentials 
$B_\mu^a$ and structure constants $f_{abc}$, this readily follows since
\be
\frac{g^2}{64\pi^2} \epsilon^{\mu\nu\lambda\sigma}  \sum_a G^a_{\mu\nu} G^a_{\lambda\sigma} = 
\frac{g^2}{16\pi^2}  \epsilon^{\mu\nu\lambda\sigma} \partial_\mu 
\Bigl( \sum_{a} B^a_\nu \partial_\lambda B^a_\sigma 
+ \frac{g}{3} \sum_{a,b,c} f_{abc} B^a_\nu B^b_\lambda B^c_\sigma \Bigr) \, .
\label{anomals}
\ee
Therefore, in perturbation theory the anomaly term drops out from the stress energy tensor. Hence 
in perturbation theory before chiral symmetry breaking, the parameter $\theta$ is a 
fixed constant, but by (\ref{anomals}), in this regime all the states
parameterized by a fixed value of $\theta \in [0,2\pi]$ with all other quantum 
numbers equal are energetically degenerate. Specifically
the Poincar\'e-invariant states with an arbitrary fixed $\theta$ are all admissible as 
the vacuum states of QCD, and, as noted, give rise to the
infinitely many degenerate superselection sectors of perturbation 
theory before chiral symmetry breaking.

Since the anomaly term is a total derivative, we can rewrite the right-hand side of (\ref{anomals}) as
\be 
\frac{g^2}{32} \epsilon^{\mu\nu\lambda\sigma} 
\sum_a G^a_{\mu\nu} G^a_{\lambda\sigma} = \partial_\mu K^\mu \, ,
\label{dualanomals}
\ee
where $K_\mu$ is the Chern-Simons current. We can trade this auxiliary composite vector field for its
Hodge dual $3$-form, $K_\mu = \epsilon_{\mu\nu\lambda\sigma} A^{\nu\lambda\sigma}/6 $, which, since
\be
\partial_\mu K^\mu = \frac16 \epsilon^{\mu\nu\lambda\sigma} \partial_\mu A_{\nu\lambda\sigma} = 
\frac{1}{24} 
\epsilon^{\mu\nu\lambda\sigma} F_{\mu\nu\lambda\sigma} \, , 
~~~~ {\rm with} ~~~~ F_{\mu\nu\lambda\sigma}
= 4  \partial_{[\mu} A_{\nu\lambda\sigma]} \, , 
\label{forms}
\ee 
yields 
\be
F_{\mu\nu\lambda\sigma} = \frac34 g^2 \sum_a G^a_{[\mu\nu} G^a_{\lambda\sigma]} \, ,
\label{4formcpterm}
\ee
with $[\ldots]$ denoting antisymmetrization over the enclosed indices. 
Thus the $\theta$-term in the perturbative Lagrangian ${\cal L}_{\tt QCD}$ can be rewritten as 
\be
\frac{1}{48 \pi^2} \theta \epsilon_{\mu\nu\lambda\sigma} F^{\mu\nu\lambda\sigma} \, .
\label{4theta}
\ee

In the presence of quarks with a mass matrix ${\cal M}$, the anomaly term (\ref{cpterm}) is interpreted as
receiving contributions from both the \`a priori ``vacuum angle" $\theta$ and the overall phase of the
quark mass matrix ${\tt Arg} \det {\cal M}$. This contribution is 
extracted from ${\cal M}$ via a change of fermion
basis implemented by chiral gauge transformations. Hence in the basis where the determinant of the
quark mass matrix is real, the total CP-violating phase is 
\be
\theta \rightarrow \theta_{\cancel{\tt CP}} = \hat \theta = \theta + {\tt Arg} \det {\cal M} \, .
\label{cptheta}
\ee
We will work in this basis in what follows. 

The formula (\ref{cptheta}) illuminates the circumstances 
which lead to CP violation very clearly. But it also points to one possible solution. 
If even one of the quarks were exactly massless, the 
determinant of ${\cal M}$ would have been zero, and the phase ${\tt Arg} \det {\cal M}$ would
have been completely arbitrary. That would have implied that the total CP-violating phase 
$\hat \theta$  is completely ambiguous, and that it could have always been chosen to be zero
by freely picking an appropriate value of ${\tt Arg} \det {\cal M}$ to cancel $\theta$. Thus there would 
have been no physical manifestation of CP violation after all. 

A physical example of this case is QCD before chiral symmetry breaking in the UV
where all the quarks are massless, and so $\hat \theta \ne 0$ is unphysical.
The problem arises after complete chiral symmetry breaking, in the IR, when the quarks are massive 
(see the still-relevant discussion in \cite{Weinberg:1978uk}). We will formulate the problem and its diagnostic 
using the effective theory of the top form, which extends (\ref{4theta}) below full chiral symmetry breaking.

To establish the dictionary we review the formalism of the theory relevant for 
the top form sector. We will use the vacuum expectation values of operators ${\cal O}$ which only depend 
on the gauge field variables as used by \cite{Luscher:1978rn} after decoupling the fermions for simplicity: we assume that they are massive, which 
absorbs the mass matrix $U(1)$ phase into $\hat \theta$, and leaves us 
with a path integral over the gauge fields $B^a_\mu = \{{\cal B}\}$. Although the quark mass 
scale is comparable to the chiral symmetry breaking scale in real QCD, 
this simplification captures the key features of the dynamics well. In this limit,
\be
\langle {\cal O} \rangle_{\hat \theta} = \frac{1}{{\cal Z}[{\hat \theta}]} 
\int \[{\cal D}{\cal B}\] {\cal O} \, e^{i S + i \hat \theta \int q(x)} \, ,
\label{expec}
\ee
where 
\be
{\cal Z}[{\hat \theta}] =  \int \[{\cal D}{\cal B}\] \, e^{i S + i \hat \theta \int q(x)} \, ,
\label{partition}
\ee
is the standard partition
function with the ``source" term $ \hat \theta \int q(x)$. Here the operator $q(x)$ is defined by 
$q(x) = \frac{g^2}{64\pi^2} \epsilon^{\mu\nu\lambda\sigma} 
\sum_a G^a_{\mu\nu}(x) G^a_{\lambda\sigma}(x)$. 
The notation $\langle \ldots \rangle_{\hat \theta}$  
represents averaging with the Euclidean path integral of the gauge theory including
the anomaly term $\propto \hat \theta$. Note that by Eqs. (\ref{dualanomals}) and (\ref{forms}), 
\be
q(x) = \frac{1}{48 \pi^2} \epsilon_{\mu\nu\lambda\sigma} F^{\mu\nu\lambda\sigma} \, .
\label{qtheta}
\ee

The Euclidean continuation of the partition function (\ref{partition}) has very specific properties which
were elucidated by Vafa and Witten \cite{Vafa:1984xg}. The function ${\cal Z}[{\hat \theta}]$ has minima at
$\hat \theta =0 + 2n\pi$, where CP is restored. This implies 
that ${\cal Z}[{\hat \theta}]$ is an even function of $\hat \theta$.
In turn, the vacuum expectation value of ${\cal O} = q(x)$ itself is \cite{Luscher:1978rn}
\be
\langle q \rangle_{\hat \theta} = i {\tt H}(\hat \theta) \, ,
\label{expq}
\ee
where by translational invariance, $\langle q(x) \rangle_{\hat \theta} = \langle q(0) \rangle_{\hat \theta} \propto 
i \partial_{\hat \theta}{\cal Z}[\hat \theta]/{\cal Z}[\hat 
\theta]$. Therefore, ${\tt H}(\hat \theta)$ is an odd function, 
${\tt H}(-\hat \theta) = -{\tt H}(\hat \theta)$  \cite{Weinberg:1978uk,Luscher:1978rn}. 

For the gauge group $SU(2)$ with a particular value of $\hat \theta$ state, L\"uscher found that 
${\tt H}(\hat \theta) = 0.078 (\mu_0 \frac{8\pi^2}{g^2})^4 (\mu_0 \rho_c)^{10/3} e^{-8\pi^2/g^2(\mu_0)} \sin(\hat \theta)$,
where $\mu_0$ is a dimensional normalization parameter, $g(\mu_0)$ the running coupling and
$\rho_c$ the IR cutoff for the instanton size. A qualitatively similar result also holds for $CP^{n-1}$ 
nonlinear $\sigma$-models in $1+1$ dimensions. Similar features are also found in more realistic setups in 
\cite{Gabadadze:1997kj,Shifman:1998if,Gabadadze:2002ff}. Because  
the vacuum angle $\hat \theta$ is  
removable by chiral transformations \underbar{above complete breaking of chiral symmetry}, where at least one 
fermion remains massless, this expectation value must be zero at scales above $\Lambda_{\tt QCD}$. 
Since $q(x) = \frac{1}{48 \pi^2} \epsilon_{\mu\nu\lambda\sigma} F^{\mu\nu\lambda\sigma}$, 
below chiral symmetry breaking
the ``secret" top form field strength is not zero when $\hat \theta$ is not zero since $q \ne 0$: 
inverting the Hodge dual, and transforming to Lorentzian 
signature variables, we find 
\be
F_{\mu\nu\lambda\sigma} = - 2\pi^2 i \epsilon_{\mu\nu\lambda\sigma} q 
= 2\pi^2 \epsilon_{\mu\nu\lambda\sigma} {\tt H}(\hat \theta) \, . \label{Fatheta}
\ee
Hence $F_{\mu\nu\lambda\sigma} \ne 0$ is a faithful diagnostic of CP violation parameterized by
$\hat \theta$. Hence restoring CP is equivalent to selecting a state in which the top form
field strength is zero. 

Further, although $F^{\mu\nu\lambda\sigma}$ starts out looking like an auxiliary term, 
it turns out that it has nontrivial 
dynamics, albeit without\footnote{This follows from the unitarity of the theory.} 
local fluctuating degrees of freedom. 
To see it, we start 
with the calculation by L\"uscher \cite{Luscher:1978rn}, who showed how the $3$-form 
potential $A_{\mu\nu\lambda}$ gains a propagator, generalizing the 
work \cite{Polyakov:1975yp,DAdda:1978dle} in $CP^{n-1}$ $\sigma$-model in two dimensions. 
In Lorenz gauge $\partial_\mu A^{\mu\nu\lambda} = 0 $, the propagator is 
given by the connected Green's function computed with
the $\hat \theta$-term in the action:
\be
\frac{1}{(3!)^2} \epsilon_{\mu\alpha\beta\gamma} \epsilon_{\nu\sigma\rho\delta}  
\langle A^{\alpha\beta\gamma}(x) A^{\sigma\rho\delta}(y)\rangle_{\hat \theta} 
= \int \frac{d^4p}{(2\pi)^4} e^{ip(x-y)} \frac{p_\mu p_\nu}{p^4} {\cal X} \, , 
\label{greens}
\ee
The residue of the momentum space Green's function 
at $p^2 \rightarrow 0$ is the topological susceptibility 
${\cal X} \simeq (\Lambda_{\tt QCD})^4$ of the theory, 
defined by the integral of the time-ordered correlator $T\bigl(q(x)q(0)\bigr)$: 
\be
{\cal X} =  - i \int d^4x \, \langle T\bigl(q(x)q(0)\bigr) \rangle_{\hat \theta} \, .
\label{suscept}
\ee
Its precise form is model-dependent. Some explicit examples 
are discussed by L\"uscher in \cite{Luscher:1978rn}.
Since the momentum space Green's function is
${\cal X} = -4\pi^4 i \, \partial_{\hat \theta} \langle q(0) \rangle_{\hat \theta} = 
4\pi^4 \partial_{\hat \theta} {\tt H}({\hat \theta})$, the function ${\tt H}$ is
\be
{\tt H}(\hat \theta) = {\cal X} \hat \theta + \ldots \, .
\label{ffun}
\ee

The full quantum effective Lagrangian includes even power terms 
$\propto \bigl( F_{\mu\nu\lambda\sigma}^2 \bigr)^{n}$,
induced by non-perturbative physics, starting with the quadratic 
\cite{DiVecchia:1980yfw,Nath:1979ik,Witten:1980sp}. They are duals to the higher-order correlators $\langle q(x_1) q(x_2) \ldots q(x_n) \rangle_{\hat \theta}$,
which are nonzero below chiral symmetry breaking/confinement scale (when ${\cal X} \ne 0$)
while vanishing when chiral symmetry is restored (and so ${\cal X} = 0$). We can reverse-engineer 
the action to restore the quadratic $4$-form 
field strength $F_{\mu\nu\lambda\sigma}$ by 
reproducing the expression for the Green's function (\ref{greens}), by 
recalling the canonical $3$-form field theory, 
which in Lorenz gauge is ${\cal L}_{{\cal F}^2} = - \frac{1}{48} {\cal F}_{\mu\nu\lambda\sigma}^2$ 
with $\partial_\mu {\cal A}^{\mu\nu\lambda} = 0$. Its momentum space Green's function is 
(see e.g. \cite{Kaloper:2016fbr}, Eq. (34) with $\xi = 0$)
\be
\langle {\cal A}_{\mu\nu\lambda}(p) {\cal A}_{\mu'\nu'\lambda'}(-p) \rangle 
=  \epsilon_{\mu\nu\lambda\rho} \epsilon_{\mu'\nu'\lambda'\rho'} \frac{p^\rho p^{\rho'}}{p^4} \, .
\label{4fprmgreens}
\ee
Comparing this expression with the Fourier transform of Eq. (\ref{greens}),
\be
\langle A_{\mu\nu\lambda}(p) A_{\mu'\nu'\lambda'}(-p) \rangle_\theta 
=  \epsilon_{\mu\nu\lambda\rho} \epsilon_{\mu'\nu'\lambda'\rho'} 
\frac{p^\rho p^{\rho'}}{p^4} 4\pi^4 {\cal X} \, ,
\label{ftgreens}
\ee
shows that the momentum space tensor structure is the same, 
and that the Green's function residue at $p^2 \rightarrow 0$
leads to the wavefunction renormalization 
\be
{\cal A}_{\mu\nu\lambda} = \frac{1}{2\pi^2 \sqrt{\cal X}} \, A_{\mu\nu\lambda} \, .
\label{wfren}
\ee
Therefore, switching to the canonically normalized $4$-forms,
the theory to quadratic order in ${\cal F}_{\mu\nu\lambda\sigma}$ includes 
\be
{\cal L}_{\tt QCD} \ni \frac{\sqrt{\cal X}}{24} 
\hat \theta \epsilon_{\mu\nu\lambda\sigma} {\cal F}^{\mu\nu\lambda\sigma}  
- \frac{1}{48} {\cal F}_{\mu\nu\lambda\sigma}^2  \, .
\label{efflag} 
\ee
up to gauge-fixing terms. 
We must highlight similar analysis given in \cite{Gabadadze:1997kj,Gabadadze:2002ff} who point out 
that the top form is an emergent 
degree of freedom below the chiral symmetry breaking scale, which encapsulates faithfully
the long range physics induced by the nonperturbative effects. 
Here, for our purposes it will suffice to keep terms only up to quadratic, ignoring higher order terms 
discussed in \cite{DiVecchia:1980yfw,Nath:1979ik,Witten:1980sp}.

As is well known, classically the $4$-form field strength does not include 
new local degrees of freedom. Briefly, by total antisymmetry the $4$-form
field strength ${\cal F}_{\mu\nu\lambda\sigma}$ is proportional to the action measure, 
${\cal F}_{\mu\nu\lambda\sigma} dx^\mu dx^\nu dx^\lambda dx^\sigma = 4! {\cal F} d^4x$,  
and then the field equation, for a purely quadratic 
theory\footnote{For a bilinear theory, the equation is corrected by a total derivative, as shown below.}, is 
$d \, {}^*{\cal F} = 0$, which yields $\partial_\mu {\cal F} = 0$, and so ${\cal F} = {\rm const}$. 
This constant of integration 
is the hidden constant elucidated in 
\cite{Aurilia:1980xj,Rosenzweig:1979ay,DiVecchia:1980yfw,Nath:1979ik,Witten:1980sp}. 

It is convenient and illuminating to canonically transform the action 
(\ref{efflag}) by trading the electric field strength ${\cal F}_{\mu\nu\lambda\sigma}$ for its magnetic dual
${\cal F}_{\mu\nu\lambda\sigma} \epsilon^{\mu\nu\lambda\sigma}/4!$. The dualization 
procedure is discussed at length in 
\cite{Aurilia:1978dw,Aurilia:1980xj, Dvali:2005an,Kaloper:2016fbr,Kaloper:2022oqv,Kaloper:2022utc}. 
To carry this out, we add a 
Lagrange multiplier term to the $4$-form action, 
recasting it in the first order form,
\be
S_{\cal F} = \int d^4 x \Bigl(- \frac{1}{48} {\cal F}_{\mu\nu\lambda\sigma}^2 
+  \frac{\sqrt{\cal X}}{24} \hat \theta \epsilon_{\mu\nu\lambda\sigma} {\cal F}^{\mu\nu\lambda\sigma}  
+ \frac{\cal F}{24}  \epsilon^{\mu\nu\lambda\sigma} 
\bigl({\cal F}_{\mu\nu\lambda\sigma} - 4 \partial_\mu {\cal A}_{\nu\lambda\sigma} \bigr) \Bigr) \, .
\label{cantra}
\ee
Next, we treat ${\cal F}, {\cal F}^{\mu\nu\lambda\sigma}, {\cal A}_{\nu\lambda\sigma}$ 
as independent variables, complete the squares, shift the
electric $4$-form variable 
$\tilde {\cal F}^{\mu\nu\lambda\sigma} = {\cal F}^{\mu\nu\lambda\sigma} 
- \bigl(\sqrt{\cal X} \hat \theta + {\cal F} \bigr) \epsilon^{\mu\nu\lambda\sigma}$
and integrate it out, using the path integral formulation of the theory. That leaves us with the dual action
\be
S_{\cal F} = \int d^4 x \Bigl(- \frac{{\cal X}}{2}  
\bigl(\hat \theta + \frac{\cal F}{\sqrt{\cal X}} \bigr)^2 - \frac{\cal F}{6}  
\epsilon^{\mu\nu\lambda\sigma}  \partial_\mu {\cal A}_{\nu\lambda\sigma} \Bigr) \, .
\label{cantrad}
\ee
Since ${\cal F}_{\mu\nu\lambda\sigma}$ is a rank-4 tensor, ${\cal F}$ is a pseudoscalar. 
Because the action (\ref{cantrad}) is a first-order formulation 
of the theory, both ${\cal F}$ and ${\cal A}_{\mu\nu\lambda}$
are independent variables to be varied over. On the other hand ${\cal X}$ 
and $\hat \theta$ are fixed parameters -- i.e. integrals of motion -- when 
we ignore additional terms that can extend (\ref{cantrad}). The 
field equations that follow are
\be
\delta {\cal A}_{\nu\lambda\sigma} : ~~~ \partial_\mu {\cal F} = 0 \, , 
~~~~~~~ \delta {\cal F}: ~~~ \epsilon^{\mu\nu\lambda\sigma} {\cal F}_{\mu\nu\lambda\sigma} 
= 4 \epsilon^{\mu\nu\lambda\sigma}\partial_{[\mu}{\cal A}_{\nu\lambda\sigma]} 
= - 4! \Bigl(\sqrt{\cal X}\hat \theta + {\cal F} \Bigr) \, .
\label{eomsc}
\ee
Note also that the quadratic term 
\be
V = \frac{{\cal X}}{2}  \bigl(\hat \theta + \frac{\cal F}{\sqrt{\cal X}} \bigr)^2 \, , 
\label{potential}
\ee
is precisely the contribution to the vacuum energy when CP is violated (see 
e.g. \cite{Gabadadze:1997kj,Gabadadze:2002ff}). 
Further note that now the total CP-violating phase is
\be
\theta_{\cancel{\tt CP}} = \hat \theta + \frac{\cal F}{\sqrt{\cal X}} \, .
\label{totphase}
\ee
The expression for $\theta_{\cancel{\tt CP}}$ becomes more intricate with the addition of new 
phases that arise with the inclusion of subleading nonperturbative corrections, which need
to be carefully collected together.

At first glance the potential term (\ref{potential}) seems puzzling when one recalls the familiar 
dilute instanton gas approximation for it,
\be
V(\theta_{\cancel{\tt CP}}) = \sum_\ell V_\ell \bigl(1- \cos(\ell \theta_{\cancel{\tt CP}})\bigr) \, ,
\label{cospot}
\ee
which is manifestly periodic in $\theta_{\cancel{\tt CP}}$. However, the magnetic dual ${\cal F}$
is a {\it variable}, which is fixed to a constant only on shell, by field
equations (\ref{eomsc}), which have infinitely many solutions. Hence,
(\ref{potential}) classically has a continuous shift
symmetry $\hat \theta \rightarrow \hat \theta (1+ \vartheta)$, 
${\cal F} \rightarrow {\cal F}- \sqrt{\cal X} \vartheta \hat \theta$,
which is broken to discrete shifts $\vartheta = 2\pi n$ by the quantization 
of ${\cal F}$, that follows from the quantization of charges in the units of $\hat \theta$.
Thus, the potential (\ref{potential}) should be understood as a combination of individual 
branches generated by ${\cal F} \rightarrow {\cal F}- 2\pi \sqrt{\cal X}$, which is the foundation
of the flux monodromy models. Of course, the 
quadratic (\ref{potential}) will be modified when higher order
top form corrections are included, but the branch structure will remain.

In retrospect, these are the exact same equations that arise if we vary the 
theory based on the Lagrangian (\ref{efflag}) with respect to ${\cal A}_{\nu\lambda\sigma}$. 
Recalling that since ${\cal F}_{\mu\nu\lambda\sigma} = 4 \partial_{[\mu}{\cal A}_{\nu\lambda\sigma]}$, 
we get the single field equation
\be
\partial_\mu \Bigl( {\cal F}^{\mu\nu\lambda\sigma} - 
\sqrt{\cal X} \hat \theta \epsilon^{\mu\nu\lambda\sigma} \Bigr) = 0 \, ,
\label{singlefield}
\ee
which after integrating once and contracting both sides of the 
integral with $\epsilon_{\mu\nu\lambda\sigma}$
yields precisely the two Eqs. (\ref{eomsc}), with ${\cal F}$ as the integration constant. 
In what follows, without any loss 
of generality, we will mainly resort to the magnetic pseudoscalar dual because the
ensuing picture is simpler. 

\section{CP Violation and CP Restoration}

\subsection{The QCD Top Form as a Diagnostic of CP Violation}

The emergent top form is a direct probe of CP violation in the theory. This is obvious even before 
the inclusion of the top form dynamics \`a la L\"uscher \cite{Luscher:1978rn}, as displayed by Eqs. 
(\ref{qtheta}), (\ref{expq}), (\ref{ffun}) and (\ref{wfren}). Combining these, 
to the leading order of the non-perturbatively corrected effective theory we find, 
in the Lorentzian signature, and taking into account our redefinition Eq. (\ref{wfren}) 
\be
\frac{\sqrt{\cal X}}{24} \epsilon_{\mu\nu\lambda\sigma} 
{\cal F}^{\mu\nu\lambda\sigma}  =  - {\cal X} \hat \theta \, .
\label{qFtheta}
\ee
Inverting the Hodge dual yields
\be
{\cal F}_{\mu\nu\lambda\sigma}  =  \sqrt{\cal X} \hat \theta \epsilon_{\mu\nu\lambda\sigma} \, .
\label{Ftheta}
\ee
As we explained explicitly in the previous section, this shows that the top form ${\cal F}_{\mu\nu\lambda\sigma} $ 
is a faithful probe of CP violation below chiral symmetry breaking:
if CP is unbroken, $\hat \theta = 0$ and so ${\cal F}_{\mu\nu\lambda\sigma} = 0$. Conversely, if 
${\cal F}_{\mu\nu\lambda\sigma} \ne 0$, $\hat \theta \ne 0$ as well and CP is violated. 

The next order of the non-perturbatively corrected 
theory becomes more interesting. The quadratic 
$\cal F_{\mu\nu\lambda\sigma}$-terms extend the 
dynamics of the top form, which now is controlled by the 
equation (\ref{singlefield}) for the electric field strength, whose solution is 
\be
{\cal F}_{\mu\nu\lambda\sigma} = \bigl(\sqrt{\cal X} \hat \theta + {\cal F} \bigr) 
\epsilon_{\mu\nu\lambda\sigma} 
= \sqrt{\cal X}  \theta_{\cancel{\tt CP}} \, \epsilon_{\mu\nu\lambda\sigma} \, ,
\label{soln4form}
\ee
as per Eq. (\ref{totphase}). This includes the totally arbitrary magnetic dual of 
top form, ${\cal F}$, which is an as-yet unspecified 
integration constant, and which arises due to the presence of the quadratic ${\cal F}^2$ in the Lagrangian (\ref{efflag}). 
I.e. such a term only appears after full chiral symmetry breaking, in the IR regime. 
One can see that this term is precisely the integration constant 
noted in \cite{Aurilia:1980xj,Rosenzweig:1979ay,DiVecchia:1980yfw}, which we reviewed above. 

Now, first off, we see that (\ref{soln4form}) is still a solution of (\ref{singlefield}), which means that
${\cal F}_{\mu\nu\lambda\sigma}$ remains a faithful probe of CP violation. Secondly, from (\ref{cptheta}),
it is clear that the integration constant ${\cal F}$, the ``pure glue" vacuum angle $\theta$ and the overall 
phase of the quark mass matrix are
completely degenerate in the full low energy theory. 
The only difference between $\theta$ and ${\cal F}$ is that ${\cal F}$ 
could be attributed to boundary conditions which select the solutions of (\ref{singlefield}). 

This degeneracy between ${\cal F}$ and $\theta$ points to a possible way of addressing the observational
fact that strong CP symmetry is  approximately valid. If the boundary conditions for (\ref{singlefield})
could be chosen so that they select ${\cal F}  = - \sqrt{\cal X} \hat \theta$, screening $\hat \theta$ with the
top form flux, the strong CP would be completely restored. Alternatively, if additional degrees of freedom
are introduced in the theory, which can dynamically adjust the flux to yield ${\cal F}  = - \sqrt{\cal X} \hat \theta$, 
then the CP symmetry would also get restored. 

\subsection{The Axion Solution}

Let us review here an explicit example of dynamical restoration 
\cite{Peccei:1977hh} of strong CP symmetry, which relies on the axion, introduced in 
\cite{Weinberg:1977ma,Wilczek:1977pj}. The axion could be any dynamical phase with anomalous QCD
couplings, that appears in the effective theory via the modification of the vacuum angle $\theta$
in  (\ref{cpterm}) by 
\be
\theta \rightarrow \theta + \frac{\phi}{f_\phi} \, \label{thetaax}
\ee
where $f_\phi$ is the axion decay constant.
In the case of PQ axion, it is controlled by the PQ 
symmetry breaking scale \cite{Peccei:1977hh}, and realistic models 
were considered in \cite{Kim:1979if,Shifman:1979if,Zhitnitsky:1980tq,Dine:1981rt}.

Once a dynamical field $\phi$ is added, the theory must be Lorentz-covariantized by adding kinetic terms for $\phi$. 
This means that the theory of the top form extended with the axion in terms of the magnetic dual 
of the top form is given by the (pseudo-)scalar field action
\be
S_{{\cal F} + \phi} = \int d^4 x \Bigl(\frac12 (\partial \phi)^2- \frac{{\cal X}}{2}  
\bigl(\frac{\phi}{f_\phi}  + \hat \theta + \frac{\cal F}{\sqrt{\cal X}} \bigr)^2 - \frac{\cal F}{6}  
\epsilon^{\mu\nu\lambda\sigma}  \partial_\mu {\cal A}_{\nu\lambda\sigma} \Bigr) \, .
\label{cantradax}
\ee
Dualizing this action back to the electric top form yields, 
after integrating out ${\cal F}$, the theory in the 
mixed top form/scalar picture is given by \cite{Dvali:2005an,Aurilia:1980jz,Kaloper:2016fbr}
\be
S_{{\cal F} + \phi} = \int d^4 x \Bigl(\frac12 (\partial \phi)^2- \frac{1}{48} 
{\cal F}_{\mu\nu\lambda\sigma}^2 
+  \frac{\sqrt{\cal X}}{24} (\frac{\phi}{f_\phi} + \hat \theta ) 
\epsilon_{\mu\nu\lambda\sigma} {\cal F}^{\mu\nu\lambda\sigma} \Bigr) \, .
\label{cantraax}
\ee
Finally, redefining the axion to $\varphi = \phi + f_\phi \hat \theta$ and defining the axion
mass $m = \sqrt{\cal X}/f_\phi$, we can eliminate the scalar $\varphi$ from the theory and rewrite
the action purely in terms of the top form and its gauge potential. 
After integrating by parts the last term,  
completing the square for $\partial \varphi$, and then integrating out $\partial \varphi$ by trading it
for the St\"uckelberg field $h_{\nu\lambda\sigma}$, we find \cite{Kaloper:2016fbr}
\be
S_{{\cal F} + \phi} = \int d^4 x \Bigl(- \frac{1}{48} {\cal F}_{\mu\nu\lambda\sigma}^2 
+  \frac{m^2 }{12} \bigl({\cal A}_{\nu\lambda\sigma} - h_{\nu\lambda\sigma} \bigr)^2 
+ \frac{m \varphi}{6}  \epsilon^{\mu\nu\lambda\sigma} \partial_\mu h_{\nu\lambda\sigma} \Bigr)\, .
\label{varcantraax}
\ee
Now $\varphi$ is an auxiliary field; 
integrating it out reduces (\ref{varcantraax}) to the massive
$4$-form gauge theory in an axial gauge. On the other hand, 
integrating out $h_{\mu\nu\lambda}$ brings the theory back to the 
mixed top form/axion picture (\ref{cantraax}). 

The actions (\ref{cantradax}), (\ref{cantraax}), (\ref{varcantraax}) all represent the same theory,
as has been discussed\footnote{In particular, these 
actions were used in the formulation of high scale monodromy inflation 
\cite{Kaloper:2008qs,Kaloper:2008fb,Kaloper:2011jz}
due to the large degree
of symmetry which they exhibit, which, once mass scale $m$ is fixed, protects the theory from both perturbative and nonperturbative
corrections in quantum field theory and in quantum gravity.} at some 
length before \cite{Dvali:2005an,Aurilia:1980jz,Kaloper:2016fbr}. In this work, our main 
interest is how the vacuum selection occurs in these descriptions, and what are its 
implications for the CP problem. The specifics were discussed in \cite{Dvali:2005an,Aurilia:1980jz},
which we follow here. The CP restoration occurs via the familiar axion rolling in the scalar axion picture
(\ref{cantradax}), where the field equations and Poincar\'e invariance of the 
vacuum of the theory yield ${\cal F} = {\rm const.}$, $\partial \phi = 0$. Therefore
the vacuum value of $\phi$ is determined precisely by
\be
\frac{\phi}{f_\phi} + \hat \theta + \frac{\cal F}{\sqrt{\cal X}} = 0 \, .
\label{axvac}
\ee
Varying the action (\ref{cantradax}) with respect to ${\cal F}$ yields
$ \epsilon^{\mu\nu\lambda\sigma} {\cal F}_{\mu\nu\lambda\sigma} 
= - 4! \sqrt{\cal X} \bigl(\phi/f_\phi + \hat \theta + {\cal F}/\sqrt{\cal X} \bigr)$.
Hence ${\cal F}_{\mu\nu\lambda\sigma}$ vanishes in this particular 
state, and so the state is CP-invariant. 

In the picture based on (\ref{cantraax}), CP restoration works similarly. Although the 
constant ${\cal F}$ does not appear in the action, it reemerges as the integration 
constant after solving the field equation for ${\cal F}_{\mu\nu\lambda\sigma}$, which yields
${\cal F}_{\mu\nu\lambda\sigma} 
= \sqrt{\cal X} \epsilon_{\mu\nu\lambda\sigma} \bigl(\phi/f_\phi + \hat \theta + {\cal F}/\sqrt{\cal X} \bigr)$.
The field equation for the axion is 
$\partial^2 \phi = \frac{m}{24}  \epsilon^{\mu\nu\lambda\sigma} {\cal F}_{\mu\nu\lambda\sigma}$, 
which in the Poincar\'e-invariant vacuum, by 
$\partial \phi = 0$, yields $\epsilon^{\mu\nu\lambda\sigma} {\cal F}_{\mu\nu\lambda\sigma} =0$.
As a consequence, the axion $\phi$ again satisfies (\ref{axvac}), and CP is again restored in this state.

Finally, in the picture based on (\ref{varcantraax}), we find just a theory of a massive
top form. In the Proca gauge, the top form field solves
\be
\partial^\mu {\cal F}_{\mu\nu\lambda\sigma} + m^2  {\cal A}_{\nu\lambda\sigma} = 0 \, .
\label{proca}
\ee
Now Poincar\'e invariance requires that $\partial^\mu {\cal F}_{\mu\nu\lambda\sigma}$ must vanish,
and then (\ref{proca}) implies that ${\cal A}_{\nu\lambda\sigma}$ must also vanish up to a 
gauge transformation. In turn,  ${\cal F}_{\mu\nu\lambda\sigma}=0$ and this state, again,
is CP invariant. 

This picture has been recognized as the Higgs phase of the top form gauge 
theory \cite{Dvali:2005an,Aurilia:1980jz} which led Dvali to infer that the axion mechanism
for restoring CP,  using the top form language, corresponds to screening of the CP violating
top form, which is induced by the gauge field mass. Actually the top form theory in the Higgs phase 
shares many features of the effective theory of superconductivity \cite{Kaloper:2016fbr}.
We note that the question about effective theory describing condensates of extended 
objects has been raised by \cite{Julia:1979ur}, and explored in \cite{Quevedo:1996uu}.

While axion is a very elegant mechanism for addressing the strong CP problem, it could have problems if
it couples to some UV physics. The new UV physics could introduce new phases in the theory,
which prevent the axion from acquiring the vacuum value (\ref{axvac}) which selects 
$\theta_{\cancel{\tt CP}} = 0$. For example, if a new potential contribution
$V(\phi)$ were to arise in (\ref{axvac}), the axion vacuum value would instead be
$m^2 f_\phi ({\phi}/{f_\phi} + \hat \theta + {\cal F}/{\sqrt{\cal X}}) + \partial_\phi V(\phi)= 0$. 
When the new potential
is steeper, being controlled by higher scales, and has minima displaced away from those in the
QCD sector, the axion vacuum would be deflected from zero to values where the total strong
CP-violating phase is 
\be
\theta_{\cancel{\tt CP}} = \frac{\phi}{f_\phi} + \hat \theta + \frac{\cal F}{\sqrt{\cal X}} = - \frac{1}{m^2 f_\phi} 
\partial_\phi V(\phi) \, ,
\label{axuv}
\ee
which would typically be greater than $10^{-10}$ without fine tuning or new dynamics in the UV. 
In such situations one may need a 
different mechanism for CP restoration.  

A clue that a different mechanism may be available is provided 
by the integration constant ${\cal F}$, which so far has been merely a spectator. 
In what follows, we will set the axion aside and focus on the general reinterpretation of
the CP violating phases as integration constants, to  
seek a different relaxation mechanisms to discharge them. 

\section{Quantal Theory of Constants}

\subsection{Top Form Charges}

It has long been argued that the $4$-form sector of the effective theory of QCD 
below chiral symmetry breaking scale must be completed with the inclusion of the 
sources charged under the top form 
\cite{Gnadig:1976pn,Aurilia:1978dw,Luscher:1978rn,Aurilia:1980xj,Gabadadze:1997kj,Gabadadze:2002ff}. 
Given the spacetime properties of the gauge field potential ${\cal A}_{\mu\nu\lambda}$, the sources 
must be codimension-1 objects, which must be compact to have finite action. 
This means that in $4D$ the top form sources are membranes. In the effective
theory we can view them as spherical domain walls separating the vacua with a different 
value of $\theta_{\cancel{\tt CP}}$, and this implies that they must have non-zero energy, 
represented by non-vanishing tension. 
 
The key motivation for introducing these objects, as argued in 
\cite{Gnadig:1976pn,Luscher:1978rn,Aurilia:1978dw,Gabadadze:1997kj,Gabadadze:2002ff}, 
is related to confinement and the existence 
of light glueball states of QCD. Since QCD confines, the QCD vacuum at 
short and long distances is believed to have very different properties, 
which can be parameterized by  ${\cal F}_{\mu\nu\lambda\sigma}$ variation over the spatial 
domain of a bound state. In particular, in the interior and exterior of the 
bound state, the value of ${\cal F}_{\mu\nu\lambda\sigma}$ is different: due to confinement, the
field ranges of gauge fields are finite, and so all of them vanish at infinity.
Inside, the theory is in deconfined phase below chiral symmetry breaking, 
and so it is characterized by ${\cal F}_{\mu\nu\lambda\sigma} \sim \theta_{\cancel{\tt CP}}$. Thus the
value of ${\cal F}_{\mu\nu\lambda\sigma}$ must be controlled by boundary conditions far away and
in the core. 

While these effects originate from strong coupling regime in a QFT, at low energies 
we can model the variation of ${\cal F}_{\mu\nu\lambda\sigma}$ 
by introducing a spherical membrane which is required to maintain Gauss' law by absorbing
the flux of ${\cal F}_{\mu\nu\lambda\sigma}$ at the boundary of the bound state, to 
confine the interior fields. We can see this explicitly by going back to 
(\ref{efflag}) and interpreting the bilinear term as a boundary term,
$\frac{\sqrt{\cal X}}{24} \hat \theta \epsilon_{\mu\nu\lambda\sigma} {\cal F}^{\mu\nu\lambda\sigma} 
= \frac{\sqrt{\cal X}}{6} \hat \theta \epsilon_{\mu\nu\lambda\sigma} \partial^\mu {\cal A}^{\nu\lambda\sigma}$,
which, after integrating over the spherical surface bounding a region 
with ${\cal F}_{\mu\nu\lambda\sigma} \ne 0$ and ${\cal F}_{\mu\nu\lambda\sigma}=0$ yields
\be
S_{\cal F} = - \int d^4 x \frac{1}{48} {\cal F}_{\mu\nu\lambda\sigma}^2 + 
\frac {\sqrt{\cal X}\hat \theta}{6} \int d^3 \xi \, {\cal A}_{\mu\nu\lambda} 
\frac{\p x^\mu}{\p \xi^a} \frac{\p x^\nu}{\p \xi^b} 
\frac{\p x^\lambda}{\p \xi^c} \epsilon^{abc} \, ,
\label{thetacharge}
\ee
where the second term involves an integral over the sphere separating
deconfined interior and an exterior vacuum with all fields confined inside a small ball.

The surface term looks precisely like a membrane charge term for an infinitely thin membrane
\cite{Gnadig:1976pn,Aurilia:1978dw}, 
\be
{\cal Q} = - \sqrt{\cal X} \hat \theta \, .
\label{qcdcharge}
\ee
To complete the membrane Lagrangian, one would also
need to add the tension term, which must arise as a measure of the rest energy of a membrane,
essentially gapping it from the vacuum \cite{Gnadig:1976pn}, which has the form
\be
- {\cal T} \int d^3 \, \xi \sqrt{|\det(\eta_{\mu\nu} \frac{\p x^\mu}{\p \xi^a} \frac{\p x^\nu}{\p \xi^b} )|} \, \, ,
\label{tension}
\ee
where the integration is over the membrane worldvolume, and ${\cal T}$ is the membrane tension. 
From this perspective, it appears that QCD in the strong coupling regime below chiral symmetry breaking
``manufactures" it's own system of membranes. This has been used to analyze the spectrum of glueballs
in QCD, which were compared to the system of membranes in strong coupling in 
interesting papers \cite{Gabadadze:1997kj,Gabadadze:2002ff}. There, and also earlier, in \cite{Gnadig:1976pn},
numerical estimates of the scales of charges and tensions of the membranes were obtained,
being set by the QCD scale $\Lambda_{\tt QCD} \simeq 100 MeV$. 

In the presence of these membranes, one would expect by comparison with 
\cite{Aurilia:1978dw,Aurilia:1980xj}, but also 
\cite{Brown:1987dd,Brown:1988kg,Kaloper:2022oqv,Kaloper:2022utc}, that the magnetic dual
top form flux ${\cal F}$ should be quantized in the units of charge ${\cal Q}$,
\be
{\cal F} = {\cal N} {\cal Q} = - {\cal N} \sqrt{\cal X} \hat \theta \, ,
\label{quantized}
\ee
where ${\cal N}$ is an integer. Further, the quantum nucleation of membranes 
\cite{Brown:1987dd,Brown:1988kg,Kaloper:2022oqv,Kaloper:2022utc}, completely analogous to the 
Schwinger processes for particle production in a background fields \cite{Schwinger:1951nm}, would
provide a channel for nonperturbative relaxation of ${\cal F}$ that could lead to a process for
adjusting $\theta_{\cancel{\tt CP}}$ in discrete steps. 

However there are questions concerning this idea, if one wants to use the membranes
``emergent" within QCD itself. First of all, it is not clear how reliable is the thin-wall approximation for the
membranes \cite{Gabadadze:1997kj}. It is conceivable that at least some of these objects are
really smeared rather than localized configurations. Specific examples were provided in the 
AdS/CFT approach to large N gauge theories in \cite{Dubovsky:2011tu}, where in fact the theory
appears to have both thin and thick ``membranes". Secondly, investigations of
the stability of the gauge theory vacua to nucleation of membranes, given in \cite{Shifman:1998if} in the large N limit, 
and in \cite{Forbes:2000et} for smaller gauge group ranks like in QCD, suggest the decay rates may be small. E.g.,
\cite{Shifman:1998if} estimates $\Gamma \sim e^{-const \times N^4}$, where for small membranes 
in approximately flat space, appropriate for this case, the constant factor in the exponent already 
includes a factor of $27\pi^2/2$ \cite{Coleman:1977py}. 
If the remaining factor were  ${\cal R} \simeq {\cal T}^4/{\cal Q}^6 \ga 1$,
on dimensional grounds, the rate would be too slow: e.g., using $N_{\tt QCD} = 3$ suggests that 
$\Gamma \sim \exp\bigl({- \frac{27 \pi^2}{2} N_{\tt QCD}^4 {\cal R}}\bigr) \la \exp\bigl({- 10^5 {\cal R}}\bigr)$. 
Another estimate \cite{Forbes:2000et} for finite $N_{\tt QCD}$ arrives at the similar 
quantitative conclusion. On the other hand, explicit examples \cite{Dvali:1998ms,Dubovsky:2011tu} exist 
which suggest that the barrier to tunneling diminishes as N grows large. 
 
This example nevertheless points to a dynamical channel that could yield
 a viable mechanism for restoring CP. Since the main unknown is the nucleation rate suppression, 
 a possible avenue is to introduce a system of ``fundamental" membranes with charges and tensions
 which are determined by dynamics outside of QCD, and then mix it with 
 the CP-violating operators in QCD. This could happen if the top form of QCD
 kinetically mixes with some fundamental $4$-form field strength which arises in the UV completion of the theory, as in the 
 supergravity examples considered in \cite{Aurilia:1980xj}, or with an emergent top form from another 
 hidden sector gauge theory. We take this bottom-up 
 route, and treat the charge and tension of the membranes as input parameters, whose values are to 
 be selected to ensure that the mechanism succeeds. 
  
  \subsection{Adding Another Top Form}

We begin by extending the magnetic dual top form action (\ref{cantrad}) with an inclusion of 
a copy of the top form sector, which is almost identical to the ``emergent" top form of QCD. The difference is that 
we also add membrane charged under the new top form, with tension ${\cal T}$ and charge ${\cal Q}$, 
which are input parameters of the theory, unrelated to the QCD scale. We ignore the 
``emergent" QCD defects reviewed above given the uncertainties about their decay rates 
derived from QCD, and our bottom-up approach. We model the nontrivial dynamics as arising 
only in the new top form sector, and continue to treat the emergent electric top form of QCD as a spectator, 
which will discharge since it mixes with the new top form. 

Our choice of the couplings of the two top forms is dictated by the axion 
couplings \cite{Weinberg:1977ma,Wilczek:1977pj}, replicating precisely 
the expression for the potential in the axion-extended action (\ref{cantradax}) 
\cite{Dvali:2005an}. Using the magnetic dual variables, the new action is
\ba
S_{{F} + {\cal H}} &=& \int d^4 x \Bigl(- \frac{1}{2}  \bigl( {\cal H} 
+\sqrt{\cal X} \hat \theta+  \frac{F}{2\pi^2 \sqrt{\cal X}} \bigr)^2 \nonumber \\
&+& \frac{1}{24\pi^4 {\cal X}} 
\epsilon^{\mu\nu\lambda\sigma}  \partial_\mu \bigl({ F }\bigr) {A}_{\nu\lambda\sigma} 
+ \frac{1}{6}  
\epsilon^{\mu\nu\lambda\sigma}  \partial_\mu \bigl( {\cal H} \bigr) {\cal C}_{\nu\lambda\sigma} 
\Bigr) \\
&-& {\cal T} \int d^3 \, \xi \sqrt{|\det(\eta_{\mu\nu} \frac{\p x^\mu}{\p \xi^a} \frac{\p x^\nu}{\p \xi^b} )|} 
- \frac{\cal Q}{6} \int d^3 \xi \, {\cal C}_{\mu\nu\lambda} \frac{\p x^\mu}{\p \xi^a} \frac{\p x^\nu}{\p \xi^b} 
\frac{\p x^\lambda}{\p \xi^c} \epsilon^{abc}  \, , \nonumber 
\label{cantradcharged}
\ea
where ${\cal H}$ is the new top form. The $3$-form ${\cal C}$ comes in as 
Lagrange multiplier enforcing the field equation $\partial {\cal H}=0$. 
Here we have reverted to using the emergent QCD top form
which is not canonically normalized, that was directly calculated by L\"uscher \cite{Luscher:1978rn}. 
The rescaling is given in Eq. (\ref{wfren}). This will aid in the phenomenological considerations to follow. 
The bilinear terms $\propto \epsilon^{\mu\nu\lambda\sigma}$ in the first row were integrated by 
parts to correctly covariantize the top forms on the membranes, because dynamically they are 
spherical boundaries of regions with different $\theta_{\cancel{\tt CP}}$ 
(see, e.g. \cite{Kaloper:2022oqv,Kaloper:2022utc}). Here ${\cal T}$ and ${\cal Q}$ are related to a 
mass scale ${\cal M}$ which controls the new top form dynamics, be it emergent or fundamental, by 
formulas like ${\cal T} = \zeta {\cal M}^3$ and ${\cal Q} = \xi {\cal M}^2$, where $\zeta$ and $\xi$ are
dimensionless constants, presumably ${\cal O}(1)$. We will confirm this expectation shortly. 

Varying the action with respect to the variables ${F}, {A}_{\nu\lambda\sigma}$ and
${\cal H}, {\cal C}_{\nu\lambda\sigma}$ yields bulk equations away from the membranes, which look
like two copies of Eqs. (\ref{eomsc}), with slightly modified terms reflecting the field contents in 
(\ref{cantradcharged}). Inverting the Hodge dual in the second of (\ref{eomsc}) and incorporating
(\ref{cantradcharged}), we obtain, using 
${F}_{\mu\nu\lambda\sigma} = 4  \partial_{[\mu} {A}_{\nu\lambda\sigma]}$, 
${\cal H}_{\mu\nu\lambda\sigma} = 4  \partial_{[\mu} {\cal C}_{\nu\lambda\sigma]}$, 
\be
\partial_\mu {F} =  
\partial_\mu {\cal H} = 0 \, , ~~~~~~~  \frac{1}{2\pi^2 \sqrt{\cal X}} {F}_{\mu\nu\lambda\sigma} 
={\cal H}_{\mu\nu\lambda\sigma} 
= \Bigl( {\cal H} + \sqrt{\cal X}  \hat \theta +  \frac{F}{2\pi^2 \sqrt{\cal X}} \Bigr) \, \epsilon_{\mu\nu\lambda\sigma} 
\, .
\label{eomsbulk}
\ee
As before, away from the membranes both fluxes ${\cal F}$ and ${\cal H}$ are constant. However, once
membranes are included, the flux ${\cal H}$ which is sourced by them will change discretely 
across a membrane, and acquire different values inside and out. 
Note that from the second of Eq. (\ref{eomsbulk}) we immediately see that after the inclusion of the second
top form ${\cal H}$, the CP-violating phase is 
\be
\theta_{\cancel{\tt CP}} =  \frac{\cal H}{\sqrt{\cal X}} + \hat \theta + \frac{F}{{2\pi^2 {\cal X}} } \, .
\label{totphases}
\ee

Note also that the theory (\ref{cantradcharged}), as it stands, is well defined after chiral symmetry breaking,
when ${\cal X} > 0$, with some subtleties to account for before chiral symmetry breaking when ${\cal X} = 0$. 
In this limit the term which depends on the vacuum angle $\hat \theta$ immediately drops out, as it should. The 
magnetic dual $F$ is decoupled; in the action  (\ref{cantradcharged}), in the limit ${\cal X} \rightarrow 0$
the topological susceptibility ${\cal X}$ plays the role analogous to $G_N \sim 1/\mpl^2$ in gravity, and so its 
vanishing implies decoupling of $F$, which is fixed to a constant background value. The only acceptable
value is zero, since otherwise the action diverges. This is analogous to how a flat space is singled out in
the weak coupling limit of gravity. Further the QCD top form hasn't emerged yet; from
Eq. (\ref{eomsbulk}) we see that $F_{\mu\nu\lambda\sigma} \equiv 0$. The CP symmetry hasn't broken,
since the theory still has massless quarks, and so the quark mass matrix phase is undetermined. 
Thus in this limit the top form part of the action 
comes down to only  
\ba
S_{{F} + {\cal H}} &\rightarrow& \int d^4 x \Bigl(- \frac{1}{2}  {\cal H}^2 + \frac{1}{6}  
\epsilon^{\mu\nu\lambda\sigma}  \partial_\mu \bigl( {\cal H} \bigr) {\cal C}_{\nu\lambda\sigma} 
\Bigr) \\
&-& {\cal T} \int d^3 \, \xi \sqrt{|\det(\eta_{\mu\nu} \frac{\p x^\mu}{\p \xi^a} \frac{\p x^\nu}{\p \xi^b} )|} 
- \frac{\cal Q}{6} \int d^3 \xi \, {\cal C}_{\mu\nu\lambda} \frac{\p x^\mu}{\p \xi^a} \frac{\p x^\nu}{\p \xi^b} 
\frac{\p x^\lambda}{\p \xi^c} \epsilon^{abc}  \, . \nonumber 
\label{topacts}
\ea
As per the discussion in Sec. 2 and in \cite{Kaloper:2022oqv,Kaloper:2022utc}, it is straightforward to see
that the first line of this action is just the dual of $ -\frac{1}{48} \int d^4x \, {\cal H}_{\mu\nu\lambda\sigma}^2$: the
top form ${\cal H}$ completely decoupled from QCD. In this limit, there may well be a term in the
vacuum angle $\theta \ni \epsilon^{\mu\nu\lambda\sigma} {\cal H}_{\mu\nu\lambda\sigma}$ which however
is invisible since it always gets compensated by another phase; it only reappears after chiral symmetry breaking.
This means, that the quantum channels which can discharge the CP-violating phase $\theta_{\cancel{\tt CP}}$
are completely inaccessible until after chiral symmetry breaking.

With this in mind, we now assume that chiral symmetry breaking has taken place, ${\cal X} > 0$, 
and we consider quantum mechanical membrane production via
Schwinger processes \cite{Schwinger:1951nm}. The original, general analysis was presented in 
\cite{Brown:1987dd,Brown:1988kg} for both theories without and with gravitational effects included. 
As we declared before, here we neglect gravity, decoupling it by 
taking the limit $\mpl \rightarrow \infty$, and consider the field theory 
limit in Euclidean space. We will confirm the validity of this approximation in what follows below. 

We need the Euclidean action to calculate the membrane 
nucleation rates \cite{Coleman:1977py,Callan:1977pt,Garriga:1993fh}.
The reformulation of (\ref{cantradcharged}) to Euclidean space is straightforward; we follow the
steps in \cite{Kaloper:2022oqv,Kaloper:2022utc}, but we ignore the gravitational terms. 
First, we Wick-rotate the action using $t = - i x^0_E$, which gives 
$- i \int d^4x \sqrt{g} {\cal L}_{\tt QFT} = - \int d^4x_E \sqrt{g} {\cal L}^E_{{\tt QFT}}$. 
Next, using 
${ A}_{0 jk} = { A}^{E}_{0jk}$, ${A}_{jkl} =  {A}^{E}_{jkl}$, and likewise for ${\cal C}$, and 
$\epsilon_{0ijk} = \epsilon^{E}_{0ijk}$ and $\epsilon^{0ijk} = -\epsilon_E^{0ijk}$, the tension and charge terms, with notation
$\gamma = |\det(\eta_{\mu\nu} \frac{\p x^\mu}{\p \xi^a} \frac{\p x^\nu}{\p \xi^b} )|$, transform to
$- i {\cal T} \int d^3 \xi \sqrt{\gamma} = - {\cal T} \int d^3 \xi_E \sqrt{\gamma}$ 
and $i {\cal Q} \int {\cal C}_i = - {\cal Q} \int {\cal C}_i$, and 
the scalars do not change. The Euclidean action, defined by $i S = - S_E$, is 
\ba
S_E&=&\int d^4x_E \Bigl( \frac{1}{2}  \bigl( {\cal H} +\sqrt{\cal X} \hat \theta+ \frac{F}{2\pi^2 \sqrt{\cal X}}   \bigr)^2 
+  \frac{1}{24\pi^4 {\cal X}} 
{\epsilon^{\mu\nu\lambda\sigma}_E} \partial_\mu \bigl( { F} \bigr) { A}^E_{\nu\lambda\sigma} 
+ \frac{1}{6} {\epsilon^{\mu\nu\lambda\sigma}_E} \partial_\mu \bigl( {\cal H} \bigr) {\cal C}^E_{\nu\lambda\sigma} \Bigr)  \nonumber \\
\label{actionnewmemeu}
&+& {\cal T} \int d^3 \xi_E \sqrt{\gamma}_{\cal C} - \frac{{\cal Q}}{6} \int d^3 \xi_E \, {\cal C}^E_{\mu\nu\lambda} \, 
\frac{\p x^\mu}{\p \xi^\alpha} \frac{\p x^\nu}{\p \xi^\beta} 
\frac{\p x^\lambda}{\p \xi^\gamma} \epsilon_E^{\alpha\beta\gamma} \, .
\ea
From here on we drop the index ${E}$. 

We can use the Euclidean action (\ref{actionnewmemeu}) to construct the configurations which are comprised of sections 
of the $4D$ sphere $S^4$, cojoined along a fixed $S^3$ lattitude sphere, which represents the Euclidean worldvolume of a 
spherical membrane bearing the tension ${\cal T}$ and charge ${\cal Q}$. Detailed constructions of such configurations
and an investigation of their properties can be found in \cite{Brown:1987dd,Brown:1988kg,Kaloper:2022oqv,Kaloper:2022utc},
where many configurations arise because they are supported by gravitational effects. When gravity is decoupled by
taking the limit $\mpl \rightarrow \infty$, only one of those 
configurations remains viable. This configuration is the Euclidean bounce 
relating the backgrounds with non-negative vacuum energy given by 
\be
V = \frac{1}{2}  \bigl( {\cal H} +\sqrt{\cal X} \hat \theta+  \frac{F}{2\pi^2 \sqrt{\cal X}} \bigr)^2 \, ,
\label{finpots}
\ee
where the interior and exterior regions have the values of ${\cal H}$ that differ from one another by a unit of charge. 
This follows immediately from (\ref{actionnewmemeu}): varying this action with respect to ${\cal C}_{\nu\lambda\sigma}$ 
and taking the induced metric on the membrane to be $S^3$ 
yields \cite{Brown:1987dd,Brown:1988kg,Kaloper:2022oqv,Kaloper:2022utc}
\be
n^\mu \partial_\mu {\cal H} = {\cal Q} \delta(r-r_0) \, , 
\label{Heq}
\ee 
where $n^\mu$ is the outward-oriented normal to the membrane, $r_0$ its radius and $r$ the coordinate along $n^\mu$. 
Integrating this gives
\be
{\cal H}_{out} - {\cal H}_{in} = {\cal Q} \, ,
\label{delH}
\ee
as claimed. Since ${\cal H}$ can only vary in steps of ${\cal Q}$, we can take 
it to be quantized, ${\cal H} = {\cal N} {\cal Q}$. However,
${\cal H}$ is degenerate with $\sqrt{\cal X} \hat \theta$ and  
$\frac{F}{2\pi^2 \sqrt{\cal X}}$, and both of these variables
can take an arbitrary value. E.g., $\hat \theta \ni {\tt Arg} \det {\cal M}$, 
which given the arbitrariness of the overall phase of the quark mass
matrix, implies that $\hat \theta$ is a continuous variable on a circle. 
Thus we will ignore the question about the quantization of ${\cal H}$ since 
it is moot here. 

There is one more constraint governing membrane dynamics, which 
follows from energy conservation. The energy difference between
the values of $V$ inside and outside the membrane, induced by the 
jump in ${\cal H}$ by a unit of membrane charge ${\cal Q}$ in Eq. (\ref{Heq}), 
must be precisely balanced by the membrane tension ${\cal T}$. The membrane nucleates with the radius
set by total energy conservation, so that the gain in energy incurred by the decay of ${\cal H}$ will be compensated by the cost 
of nucleating a membrane whose total rest energy is given by tension times its proper Euclidean volume \cite{Coleman:1977py}. 
To find it, starting with the action (\ref{actionnewmemeu}), 
we note that (\ref{delH}) implies that the bilinear terms from the top line of  (\ref{actionnewmemeu}) cancel agains the charge term 
from the bottom line of (\ref{actionnewmemeu}). Next we recall that the relevant action governing the nucleation dynamics is the 
tunneling bounce action, which is a difference of the action  (\ref{actionnewmemeu}) on a geometry which includes a single bubble, 
and the same action on the initial smooth background geometry without any bubbles. Subtracting the two, 
what remains is the gain minus the cost: the integral of the potential difference over the volume of $S^4$, minus 
the tension integrated over the membrane volume $S^3$ \cite{Coleman:1977py}. 
Since both integrands are constant over their respective domain of integration, and the relevant 
volume factors are $V_{S^4} = \pi^2 r_0^4/2$ and $V_{S^3} = 2\pi^2 r_0^3$, formally the membrane radius
minimizes the action 
\be
S_{membrane} = 2\pi^2 r_0^3 {\cal T} - \frac12 \pi^2 r_0^4 \Delta V \, ,
\label{sbounce}
\ee
where $\Delta V= V_{out}({\cal H}_{out}) - V_{in}({\cal H}_{in})$, and ${\cal H}_{out,in}$ are related to each other by (\ref{delH}). 
Minimizing (\ref{sbounce}) with respect to $r_0$ yields the membrane radius at nucleation \cite{Coleman:1977py},
\be
r_0 = \frac{3{\cal T}}{\Delta V } \, , 
\label{radnuc}
\ee
and, by evaluating (\ref{sbounce}) at $r=r_0$, we find the actual bounce action \cite{Coleman:1977py}
\be
B = \frac{27\pi^2}{2} \frac{{\cal T}^4}{\bigl(\Delta V\bigr)^3} \, .
\label{bounce}
\ee

The bubble nucleation rate per unit time per unit volume then is given by 
$\Gamma = A e^{-B}$ \cite{Coleman:1977py,Callan:1977pt}, and to compute it we need
the prefactor $A$. The prefactor for membrane production in the flat space limit, 
where gravity is negligible, has been obtained by Garriga in 
\cite{Garriga:1993fh}, and using his results, we finally obtain
\be
\Gamma \simeq 9 \frac{{\cal T}^4}{\bigl(\Delta V\bigr)^2} 
\exp\Bigl({-\frac{27\pi^2}{2} \frac{{\cal T}^4}{\bigl(\Delta V\bigr)^3}}\Bigr) \, .
\label{nucrate}
\ee

Note that inside each membrane, the potential $V$ jumps by approximately
\be
\Delta V =  {\cal Q}  \Bigl( {\cal H} +\sqrt{\cal X} \hat \theta+  \frac{F}{2\pi^2 \sqrt{\cal X}} \Bigr) \, ,
\ee
The top form also decreases because it depends on ${\cal H}$. Inside each new bubble, the top form
decreases by
\be
\Delta { F}_{\mu\nu\lambda\sigma} = {2\pi^2 \sqrt{\cal X}}  {\cal Q} \, \epsilon_{\mu\nu\lambda\sigma} \, .
\label{qcdtopchange}
\ee
Since the membrane tension is strictly positive, membrane discharge 
drives $\theta_{\cancel{\tt CP}}$ toward restoring CP, as dictated by \cite{Vafa:1984xg}. The reverse processes 
are excluded by (\ref{radnuc}) in the flat space limit, since transitions increasing the potential and making
CP violation worse require $\Delta V <0$, which with positive membrane tension also requires $r_0 <0$, which is unphysical. 

An extension of this argument shows that discharges 
must stop before $F_{\mu\nu\lambda\sigma} = 0$, unless the initial 
value of $F_{\mu\nu\lambda\sigma}$ is fine tuned to be an 
integer multiple of ${2\pi^2 \sqrt{\cal X}} {\cal Q}$. 
Once $| F_{\mu\nu\lambda\sigma} | <{2\pi^2 \sqrt{\cal X}}  {\cal Q}$, 
the next discharge would have to produce a state with 
$| F_{\mu\nu\lambda\sigma} | >{2\pi^2 \sqrt{\cal X}}  {\cal Q}$, which would 
{\it increase} the potential inside the membrane. But for such a 
bubble to get on shell, $r_0$ would have to be negative by Eq. (\ref{radnuc}), 
since ${\cal T} > 0$ and $\Delta V<0$. Again, this is unphysical; therefore the
nucleation processes terminate when $| F_{\mu\nu\lambda\sigma} | < {2\pi^2 \sqrt{\cal X}} {\cal Q}$. 
This was also found in \cite{Brown:1988kg}, and it remains true regardless of whether the discharges 
are applied to relaxing the CP-violating phase or the cosmological constant. 

Further conditions on the dynamics come from phenomenological considerations. The discharge cascade needs to 
relax $\theta_{\cancel{\tt CP}}$ from an initial natural value of ${\cal O}(1)$ down to $\theta_{\cancel{\tt CP}} \la 10^{-10}$. 
Since $F_{\mu\nu\lambda\sigma} = {2\pi^2 {\cal X}} \theta_{\cancel{\tt CP}} \, \epsilon_{\mu\nu\lambda\sigma}$, 
the bound $| F_{\mu\nu\lambda\sigma} | < {2\pi^2 \sqrt{\cal X}}  {\cal Q}$ 
and Eq. (\ref{qcdtopchange}) means that we must take
\be
 \frac{{\cal Q}}{\sqrt{\cal X}} \la 10^{-10} \, .
\label{qbound} 
\ee
Since ${\cal X}^{1/4} \simeq {\rm few} \times 100 MeV$, this implies 
\be
{\cal Q}^{1/2} \la \sqrt{10^{-11} GeV^2} \simeq 3 keV  \, . 
\label{Qvalue}
\ee
Next, the discharges should be fast \cite{Guth:1982pn,Turner:1992tz,Freese:2004vs}. 
The nucleation rate (\ref{nucrate}) 
should be more efficient than the cosmic dilution at the time
of chiral symmetry breaking. The corresponding Hubble scale is roughly 
$H_{\tt QCD} \simeq (100 MeV)^2/\mpl \simeq 10^{-20} GeV$,
and imposing ${\Gamma}/{H_{\tt QCD}^4} \ga 1$ \cite{Guth:1982pn,Turner:1992tz,Freese:2004vs} gives 
\be
10^{81} \frac{{\cal T}^4}{\bigl(\Delta V\bigr)^2(GeV)^4} \ga \exp\Bigl({\frac{27\pi^2}{2} \frac{{\cal T}^4}{\bigl(\Delta V\bigr)^3}}\Bigr) \, .
\label{boundrate}
\ee
 The nucelation processes are the slowest close to the CP-restoring state, 
 at which point, for charges satisfying (\ref{qbound}), we have
$\Delta V = \bigl( {\cal Q} \bigr)^2 \la 10^{-22} GeV^4$. Thus we need 
\be
10^{31}\frac{{\cal T}}{(GeV)^3} \ga  \exp\Bigl({\frac{27\pi^2}{8} \times 10^{66} \times \bigl(\frac{{\cal T}}{GeV^3}\bigr)^4}\Bigr) \, .
\label{bbbound}
\ee
The right hand side term is the controlling term here. The discharges would be extremely slow unless
${\frac{27\pi^2}{8} \times 10^{66} \times \bigl(\frac{{\cal T}}{GeV^3}\bigr)^4} \la 1$. In turn this requires that 
${\cal T}^{1/3} \la 10^{-3} {MeV} = keV$, give or take. Taking into 
account the factor of $10^{31}$ on the left hand side, the numbers are a bit 
relaxed, yielding 
\be
{\cal T}^{1/3} \la 3 keV \, .
\label{tensnum}
\ee
Given (\ref{qbound}), as long as the membrane tension satisfies this inequality, the nucleation processes
would be fast enough to discharge $\theta_{\cancel{\tt CP}}$ 
within ${\cal O}(10)$ Hubble times at the QCD phase transition scale, 
even if the initial state was close to the minimal achievable $\theta_{\cancel{\tt CP}}$. 
Curiously, both ${\cal Q}$ and ${\cal T}$ appear to be set by the same scale, ${\cal M} \sim 3 keV$.

For a generic initial state with $\theta_{\cancel{\tt CP}} \gg 10^{-10}$, 
the discharges would be even faster. Indeed, Eq. (\ref{bbbound}) for a general initial 
$\theta_{\cancel{\tt CP}}$ is modified to 
\be
10^{31}\frac{{\cal T}}{(GeV)^3} \sqrt{\frac{\Delta \theta_{\cancel{\tt CP}}}{\theta_{\cancel{\tt CP}}}} \ga  
\exp\Bigl({\frac{27\pi^2}{8} \times 10^{66} \times \bigl(\frac{{\cal T}}{GeV^3}\bigr)^4} \times 
\bigl( \frac{\Delta \theta_{\cancel{\tt CP}}}{\theta_{\cancel{\tt CP}}} \bigr)^3 \Bigr) \, ,
\label{bbboundgen}
\ee
and so as $\theta_{\cancel{\tt CP}}$ increases from $10^{-10}$ toward unity, 
the left-hand side decreases as a square-root of
$\theta_{\cancel{\tt CP}}$, but the right hand side decreases much faster, as its cubic power. 
When Eq. (\ref{tensnum}) holds and $\theta_{\cancel{\tt CP}} \simeq 1$, early on in the discharge
cascade, the right-hand side is unity, for all practical intents and purposes, and the tunneling is 
extremely rapid. 

The Eq. (\ref{qbound}) implies that there would be many individual steps in the discharge cascade of 
$10^{10}$ units of $ {\cal Q}$ required to reduce $\theta_{\cancel{\tt CP}}$ 
to meet the experimental bounds. Since the nucleation processes are
so fast, this appears viable. As a comparison, the axion also needs to 
traverse -- continuously -- the same target space distance as it 
sets the modulus $\theta_{\cancel{\tt CP}}$ to zero. In any case 
as long as the tension and the charge of the membranes satisfy 
Eqs. (\ref{tensnum}) and (\ref{qbound}), respectively,
no part of the universe would end up being stuck in regions where 
$\theta_{\cancel{\tt CP}} > 10^{-10}$ for long enough to raise the question of 
why we do not see a larger CP violation. 

Since at this point we have at least some quantitative 
ideas about the scales controlling the membrane and ${\cal H}$ dynamics, we can verify 
that our neglecting gravity is justified. Let us compare the 
radius of a bubble at nucleation (\ref{radnuc}) to the Hubble length 
at the QCD chiral symmetry breaking scale. Since $H_{\tt QCD} \simeq \sqrt{\cal X}/\mpl$, we find
\be
r_0 H_{\tt QCD} \simeq \frac{\cal T}{\Delta V} \frac{\sqrt{\cal X}}{\mpl} \simeq \frac{\cal T}{\sqrt{\cal X}\mpl} 
\frac{1}{\theta_{\cancel{\tt CP} }\Delta{\theta_{\cancel{\tt CP}}}} \, .
\label{radiusbub}
\ee
Next, using Eq. (\ref{tensnum}), we find that for 
$\theta_{\cancel{\tt CP}} \sim \Delta {\theta_{\cancel{\tt CP}}} \sim 10^{-10}$, 
\be
r_0 H_{\tt QCD} \la  10^{20} \frac{\cal T}{\sqrt{\cal X}\mpl} \simeq 10^{-12} \, . 
\label{bubblesizeqcd}
\ee
Initially, when $\theta_{\cancel{\tt CP}} \sim 1$, 
the bubbles are much smaller, with $r_0 H_{\tt QCD} \la 10^{-22}$. 
So at the QCD chiral symmetry breaking scale, all the bubbles are very small compared to the background 
curvature scale $1/H_{\tt QCD} \simeq \mpl/\sqrt{\cal X} \simeq 10^{11} (eV)^{-1} \simeq 100 \, km$. 
We can always pick a large freely falling local frame, smaller than $1/H_{\tt QCD}$ and much 
larger than the size of a nucleating bubble, inside which the nucleation occurs in the same way 
as in flat space. Further, while in the presence of gravity additional tunneling processes become
possible, when the principal channel that remains as $\mpl \rightarrow \infty$ is so fast,
these new channels are inconsequential, since they are very suppressed. 
Ergo, we can ignore any gravitational corrections in the semiclassical approximation; 
a quick check is to see that the bubble nucleation equations are precisely the same as those of 
\cite{Brown:1987dd,Brown:1988kg,Kaloper:2022oqv,Kaloper:2022utc} in the limit $\mpl \rightarrow \infty$. 

We note that even if some domain walls had not completely disappeared, 
but instead had survived to the present day, or had been recreated at some scale 
much lower than the QCD chiral symmetry breaking scale, they would have remained very stealthy so far, 
provided that they do not move too fast. Since 
the tension which satisfies the bound (\ref{bbbound}) is well 
below the limits discussed in \cite{Zeldovich:1974uw}, any lingering 
domain wall fragments couldn't have affected late cosmology (i.e., induced anisotropy in the CMB) in 
too dramatic a way. A more precise reevaluation of this issue might be warranted. 
We'll set this question aside for the moment, 
and move on to a more detailed discussion of the cosmology of CP restoration.

\section{Cosmological CP Restoration} 

We assume that the universe underwent inflation followed by efficient 
reheating. After reheating the universe is evolving as a radiation-dominated, spatially flat, 
isotropic and homogeneous FRW cosmology, with temperature 
well above the QCD scale $\Lambda_{\tt QCD} \sim GeV$. 
We take this background as given, and ignore the details about where it 
came from, how the cosmological constant was cancelled, and, once we 
have the FRW metric, even the dynamical effects of gravity. 

When the early universe temperature significantly 
exceeds $\Lambda_{\tt QCD}$, the QCD sector will initially have unbroken chiral symmetry, and 
the quarks will be massless. The unbroken symmetry implies two important 
features of the dynamics: 1) initially, all the 
$\theta_{\cancel{\tt CP}}$ states will be degenerate, and 2) due to massless quarks, the mass matrix
phase ${\tt Arg} \det {\cal M}$ in Eq. (\ref{cptheta}) will be 
completely indeterminate. As a result, the QCD sector will not 
contribute to CP violation initially. 

As the universe cools, the situation changes dramatically. First, electroweak symmetry will break
as the temperature drops to below $TeV$. Subsequently chiral symmetry breaking will also occur. 
The main electroweak breaking consequence will be fixing the contribution of the
quark mass matrix to  $\theta_{\cancel{\tt CP}}$, although different $\theta_{\cancel{\tt CP}}$ 
phases may still remain degenerate until 
chiral symmetry breaking when the instanton effects arise (see, e.g. \cite{Kim:1986ax} for a review). 

Note that from Eq. (\ref{boundrate}), the discharges of $\theta_{\cancel{\tt CP}}$ induced 
by the discharges of the flux ${\cal H}$ do not occur until 
$H \la H_{\tt QCD}$ because the vacua with different $\theta_{\cancel{\tt CP}}$ are still degenerate. The 
topological susceptibility is zero early on, and the QCD top form is effectively set to zero, and so it is decoupled from 
${\cal H}$. The flux of ${\cal H}$ may change, but even its discharges not affecting  $\theta_{\cancel{\tt CP}}$ 
will be slower early
on, because thanks to our choice of the membrane tension and charge, 
the barrier to membrane nucleation will suppress more efficiently the flux decay 
for earlier times $t < 1/H_{\tt QCD}$.

As the universe cools, the electroweak breaking fixes the quark matrix and its contribution 
to the CP-breaking phase $\theta_{\cancel{\tt CP}}$. Soon after, 
the instanton contributions lift degeneracy between different $\theta_{\cancel{\tt CP}}$ states and, 
with the tunneling barrier 
becoming low, rapid discharges begin. Indeed, even if the 
initial phase varied randomly from one causal domain at 
the QCD scale to another\footnote{One would not expect this randomness 
of $\theta_{\cancel{\tt CP}}$ in a region which originates from a
single causally connected domain before inflation.}, 
these regions would develop vacuum energy induced by the condensate, 
$V = \frac12 {\cal X} \theta_{\cancel{\tt CP}}^2$, as given in Eq. (\ref{finpots}). Consequently, the domains with 
$\theta_{\cancel{\tt CP}} =  \frac{\cal H}{\sqrt{\cal X}} + \hat \theta + \frac{F}{{2\pi^2 {\cal X}} } \gg 0$ 
become unstable to nucleation of bubbles surrounded by membranes with tension ${\cal T}$ and charge ${\cal Q}$. 
Since the bubbles are very small 
compared to the background curvature, $r_0 H_{\tt QCD} \sim 10^{-22} - 10^{-12} \ll 1$, the flat 
space analysis suffices\footnote{As is commonly done, we fine-tune the cosmological constant of the 
minimal $\theta_{\cancel{\tt CP}}$ state to a sufficiently small value.} \cite{Berezin:1987bc}. 

We can see this manifestly by noting that we can use the strong CP-violating phase 
$\theta_{\cancel{\tt CP}}$ as the discharging variable. In this `axial gauge', since ${\cal H}$ is the only
variable flux, we can treat the definition of the total CP-violating phase
$\theta_{\cancel{\tt CP}} =  \frac{\cal H}{\sqrt{\cal X}} + \hat \theta + \frac{F}{{2\pi^2 {\cal X}} }$
as the change of variables from ${\cal H}$ to $\theta_{\cancel{\tt CP}}$. Adopting this viewpoint, 
we can trade ${\cal H}$ in Eq. (\ref{cantradcharged}) for $\theta_{\cancel{\tt CP}}$, and drop the 
remaining terms $\propto F$ since they do not change. In those variables, the truncated action 
(\ref{cantradcharged}) becomes, after rescaling 
${\cal C}_{\nu\lambda\sigma} = \frac{1}{\sqrt{\cal X}} {\cal Y}_{\nu\lambda\sigma}$,
\ba
S_{{\cancel{\tt CP}}} &=& \int d^4 x \Bigl(- \frac{\cal X}{2} \theta_{\cancel{\tt CP}}^2 
+ \frac{1}{6}  
\epsilon^{\mu\nu\lambda\sigma}  \partial_\mu \bigl( { \theta_{\cancel{\tt CP}}} \bigr) 
{\cal Y}_{\nu\lambda\sigma} 
\Bigr) \\
&-& {\cal T} \int d^3 \, \xi \sqrt{|\det(\eta_{\mu\nu} \frac{\p x^\mu}{\p \xi^a} \frac{\p x^\nu}{\p \xi^b} )|} 
- \frac{\cal Q}{6 \sqrt{\cal X}} \int d^3 \xi \, {\cal Y}_{\mu\nu\lambda} \frac{\p x^\mu}{\p \xi^a} \frac{\p x^\nu}{\p \xi^b} 
\frac{\p x^\lambda}{\p \xi^c} \epsilon^{abc}  \, . \nonumber 
\label{truncatedCP}
\ea
It is now obvious that the membrane discharges relax ${ \theta_{\cancel{\tt CP}}}$ in finite
steps of $\frac{\cal Q}{\sqrt{\cal X}}$. Further it is also obvious that as ${\cal X} \rightarrow 0$,
and so chiral symmetry is restored, the discharge processes are inaccessible. The 
CP-violating  contribution to the potential vanishes, and the effective charge $\frac{\cal Q}{\sqrt{\cal X}}$ for
${ \theta_{\cancel{\tt CP}}}$ diverges in this limit. This makes the barrier to tunneling 
impassable, as we can readily check by changing variables in the decay rate formula (\ref{nucrate}).

Since the potential difference between adjacent vacua is 
$\Delta V \simeq {\cal X} { \theta_{\cancel{\tt CP}}} \Delta { \theta_{\cancel{\tt CP}}}
=  \sqrt{\cal X} \cal{Q} \, { \theta_{\cancel{\tt CP}}}$,
\be
\Gamma \simeq 9 \frac{{\cal T}^4}{{\cal X} {\cal Q}^2 (\theta_{\cancel{\tt CP}})^2 } 
\exp\Bigl({-\frac{27\pi^2}{2} \frac{{\cal T}^4}{{\cal X}^{3/2} \,{\cal Q}^3 (\theta_{\cancel{\tt CP}})^3}}\Bigr) \, .
\label{nucratecp}
\ee
Recalling that  ${\cal T} = \zeta {\cal M}^3$ and ${\cal Q} = \xi {\cal M}^2$, with $\zeta$ and $\xi$ being
${\cal O}(1)$, and that ${\cal X} \simeq \Lambda_{\tt QCD}^4$, we can rewrite the bounce action in Eq. (\ref{nucratecp}) as
\be
B = \frac{27\pi^2 \zeta^4}{2 \xi^3} 
\Bigl(\frac{{\cal M}}{\Lambda_{\tt QCD}}\Bigr)^6
\frac{1}{(\theta_{\cancel{\tt CP}})^3} \, ,
\label{bth}
\ee
and the prefactor, in the units of $H_{\tt QCD}$, as
\be
\frac{A}{H_{\tt QCD}^4} = \frac{9 \zeta^4}{ \xi^2} 
\Bigl(\frac{{\cal M}}{\Lambda_{\tt QCD}}\Bigr)^8 \Bigl(\frac{\mpl}{\Lambda_{\tt QCD}}\Bigr)^4
\frac{1}{(\theta_{\cancel{\tt CP}})^2} \, .
\label{ath}
\ee
As we have chosen ${\cal M} \sim keV$, the ratios of scales are
${\cal M}/\Lambda_{\tt QCD} \sim 10^{-5}$ and $\mpl/\Lambda_{\tt QCD} \sim 2 \times 10^{19}$.
Thus, 
\be
B \simeq 100 \frac{\zeta^4}{\xi^3} \frac{1}{(10^{10} \theta_{\cancel{\tt CP}})^{3}} \, , ~~~~~~~~~~ 
\frac{A}{H_{\tt QCD}^4} \simeq \frac{\zeta^4}{\xi^2} \frac{1}{(\theta_{\cancel{\tt CP}})^2} \, ,
\label{numers}
\ee
which indeed confirm that the decay rates for the membranes with tension and charge 
in the $keV$ range are just right to restore CP. Note in particular that as soon as 
chiral symmetry breaking condensate emerges, the barrier to tunneling that leads
to membrane discharges is completely irrelevant when $\theta_{\cancel{\tt CP}} \sim 1$,
and those discharges are very fast. Only as $\theta_{\cancel{\tt CP}}$
is discharged down to $\theta_{\cancel{\tt CP}} \sim 10^{-10}$, the tunneling barrier
can suppress discharges. Further, we also see from 
Eqs. (\ref{bth}) and (\ref{ath}) that in the limit $\Lambda_{\tt QCD} \rightarrow 0$, of vanishing
chiral symmetry-breaking condensate, the membrane discharge stops. The prefactor normalized to the
QCD scale does diverge as $1/\Lambda_{\tt QCD}^{12}$, but the exponent of the bounce action converges to
zero faster.   

The discharge dynamics has some similarities with the irrational axion proposal of \cite{Banks:1991mb}. 
Since the $\theta_{\cancel{\tt CP}}\ne 0$ vacua are unstable, they will decay toward the CP-invariant 
vacuum $\theta_{\cancel{\tt CP}}=0$ by discharging the ${\cal H}$ flux. This will proceed  
as long as the emergent top form $F_{\mu\nu\lambda\sigma}$ is nonzero, and is large enough such that 
discharge reduces the potential (\ref{finpots}). These two criteria mean that the 
discharges will proceed until $| F_{\mu\nu\lambda\sigma} | <{2\pi^2 \sqrt{\cal X}}  {\cal Q}$, as we explained in Section 4.2. 
Once this value is reached, the total CP-violating phase is reduced to
\be
\theta_{\cancel{\tt CP}} \la  \frac{\cal Q}{\sqrt{\cal X}} \, .
\label{finalphase}
\ee
The nucleations will not stop any sooner than this. 

There are however important differences between our proposal and the model presented in 
\cite{Banks:1991mb}. In the irrational axion model, while the transitions between subsequent $\theta_{\cancel{\tt CP}}$
in a cascade toward $\theta_{\cancel{\tt CP}}=0$ are discrete, the step $\Delta \theta_{\cancel{\tt CP}}$
is controlled by the approximation of an integer by an irrational number, and can be arbitrarily small.
Hence, the universe in \cite{Banks:1991mb} will continue to evolve toward $\theta_{\cancel{\tt CP}}=0$.
However, the adjacent $\theta_{\cancel{\tt CP}}$ states are generically separated by huge field space distances,
and therefore the potential barrier separating them is typically very broad. Those transitions would behave
as membranes with large tension, whose nucleation rate would be extremely slow. Thus while the universe
continues to search for the state which restores CP, finding it typically would take a time much larger 
than the age of the universe. 
In contrast, in our case we consider membranes with small finite charges $ {{\cal Q}}/{\sqrt{\cal X}} \la 10^{-10}$
(\ref{qbound}) and small tension (\ref{tensnum}). For these reasons, the discharge stops at $\theta_{\cancel{\tt CP}} \la 10^{-10}$,
and this occurs fast. 

We should stress that the terminal $\theta_{\cancel{\tt CP}} \sim 10^{-10}$ should not be viewed as a prediction. 
In our proposal $10^{-10}$ is really an input, which we adopt since were are interested in a proof-of-principle.
However, if we have a $\sim keV$ scale top form and membranes, we will get that the decay rate (\ref{nucrate}) is fast compared
to the age of the universe at the chiral symmetry breaking scale, as seen in (\ref{boundrate}). Thus the nucleations would 
rapidly minimize CP-breaking effects throughout the universe by the production of bubbles of the smallest attainable
$\theta_{\cancel{\tt CP}}$. This argument actually illustrates some flexibility of our mechanism being able to address
even smaller terminal values of the strong CP phase $\theta_{\cancel{\tt CP}}$, 
if they are eventually found. For example, values of $\theta_{\cancel{\tt CP}}$ smaller than $10^{-10}$ 
could be tested by future experiments \cite{Anastassopoulos:2015ura,Zhevlakov:2020bvr}, that might reveal 
$\theta_{\cancel{\tt CP}}$ to be as small as $10^{-13}$. Even so, our mechanism would remain
operational if the charge and tension were reduced down to about 
$\sim {\cal O}(100) \, eV$. 

After nucleation, the membranes would expand at the speed of light. 
Upon their collision, they would burn away by the production 
of strongly interacting particles, i.e. predominantly pions. 
We can deduce this as follows; from the potential (\ref{finpots}) 
given above, we see that below the chiral symmetry breaking, the theory contains a term
\be
\int d^4x {\cal L}_{F + {\cal H}} \ni \frac{1}{2\pi^2 \sqrt{\cal X}} \int d^4x \, 
 {\cal H} \epsilon^{\mu\nu\lambda\sigma} F_{\mu\nu\lambda\sigma} \sim 
\frac{12}{\pi^2 \sqrt{\cal X}} \int d^4x \,  {\cal H} \partial_\mu K^\mu \, ,
\label{pions}
\ee
where in the last step we used Eqs. (\ref{dualanomals}) and (\ref{4formcpterm}). After a chiral transformation the Chern-Simons
current will shift, on shell, to $\partial K \rightarrow \partial K + f_\pi \partial^2 {\cal P}$, with ${\cal P}$ denoting the pion field. 
Substituting this into (\ref{pions}) yields 
\be
\frac{12  f_\pi}{\pi^2 \sqrt{\cal X}} \int d^4x \,  {\cal H} \partial^2 {\cal P} \, . 
\label{pionsint}
\ee
Integrating this term by parts and recalling that $\partial {\cal H} = 0$ away from membranes, and  $\Delta  {\cal H} = {\cal Q}$ 
on a membrane, we finally find 
\ba
\frac{12  f_\pi}{\pi^2 \sqrt{\cal X}} \int d^4x \,  {\cal H} \partial^2 {\cal P} &\sim&  
\frac{2  f_\pi}{\pi^2} \frac{  {\cal Q}}{\sqrt{\cal X}} \int d^3 \xi \, n \cdot \partial {\cal P} \, 
\det|\frac{\p x^\mu}{\p \xi^a} \frac{\p x^\nu}{\p \xi^b}| \nonumber \\
&\sim& 10^{-11}  f_\pi \int_{R \times S^2} d^3 \xi \, n \cdot \partial {\cal P} \, ,
\label{pionsmem}
\ea
where $n$ is the outward normal to the membrane. In this equation 
we used the bound on ${\cal Q}$ from Eq. (\ref{qbound}). 
Hence a membrane colliding with another membrane will 
bremsstrahlung pions, breaking up and transferring its energy to the pion shower. 

Although the formula  (\ref{pionsmem}) includes a suppression of $10^{-11}$, 
when the membranes are nucleated, they are
very small, $r_0 H_{\tt QCD} \sim 10^{-22} - 10^{-12}$. Since they are moving at the speed of light, by the time they 
reach the size of the horizon at the chiral symmetry breaking scale, they will have propagated through the 
background FRW for a time which is at least a factor of $10^{12}$ larger than 
their natural size. So even if they only emit a pion with a small probability 
over a period $r_0$, the emission rate will accumulate over the time of 
order the age of the universe at the QCD scale, 
and the energy transfer from the membrane walls can be completed. 

Additionally, since the relaxation dynamics occurs in discrete steps, 
and the bubbles are colliding with each other, these processes could 
produce a contribution to the stochastic gravity wave background similar to the mechanisms of 
\cite{Kosowsky:1992rz,Kamionkowski:1993fg}. Estimating the qualitative 
features of their spectrum is complicated because the bubble nucleation processes 
here are very fast and very prolific. The bubbles nucleated are very small, 
and as many as $10^{10}$ could be produced inside each Hubble domain, prompting 
a careful exploration along the lines of \cite{Kamionkowski:1993fg} 
and subsequent detailed work. For this reason
we set aside the precise discussion of gravity wave production for a later time. 

Let us mention one more possible class of phenomena which could be viewed as the signatures of the model.
It has been noted that the bubbles of other phases of QCD could be produced in high energy collisions 
\cite{Dvali:2005zk,Kharzeev:1998kz,Halperin:1998gx,Kharzeev:1998kya,Buckley:1999mv,Kharzeev:2007tn}. In 
the interior of these bubbles, one would find the CP-violating vacua, as argued in 
\cite{Kharzeev:1998kz,Halperin:1998gx,Kharzeev:1998kya,Buckley:1999mv,Kharzeev:2007tn}. In our case, 
the smaller bubbles, that would be more accessible to local scattering processes, and that would have large 
CP violation inside, would be very short-lived. Their
radius, by using Eq. (\ref{radiusbub}), would be as small as 
$r_0 \sim 10^{-14} mm$, and so we expect they would implode in about 
$\sim 10^{-26} sec$, since the interior CP-breaking phase is unstable. The large bubbles with very small CP violation 
could be longer-lived, collapsing within $\sim 10^{-16} sec$. However since already in standard QCD
similar  $\theta_{\cancel{\tt CP}}$-inducing effects can appear 
\cite{Kharzeev:1998kz,Halperin:1998gx,Kharzeev:1998kya,Buckley:1999mv,Kharzeev:2007tn}, 
given that the charge of membranes sourcing ${\cal H}$ 
is quite small, the new effects would be a correction on top of the standard 
QCD phenomena. In any case, understanding better such phenomena 
might still be interesting, and we hope to return to it in the future.

\section{Summary} 

Nonabelian gauge theories with chiral symmetry breaking have a rich and complex structure, including 
nontrivial topological sectors. Those can give rise to the non-perturbatively induced corrections to the
effective theory below the chiral symmetry breaking, which, while not
propagating local fluctuations, can mediate transitions between different topological sectors. 

An example
is a top form in QCD. It's flux is sourced by the CP-breaking phase, and so it is a faithful diagnostic of
strong CP violation. The top form flux however need not be globally constant because it could be relaxed 
by tunneling between different topological sectors in the theory. The tunneling dynamics could be 
modeled by including charged tensional membranes, which can discharge the flux by the Schwinger
processes catalyzed by the background field \cite{Schwinger:1951nm}. This mechanism is as ubiquitous 
in gauge theory as is particle pair production due to the uncertainty principle, where the pair gets on 
shell due to the work exerted by the background fields. 

Hence the discharge is not only possible, but inevitable, as long as the field energy density
exceeds the membrane rest energy. Thus a discharge of the QCD top form could be a channel for
reducing CP violation, since the top form flux is proportional to the CP-violating phase $\theta_{\cancel{\tt CP}}$, if 
the discharges are sufficiently fast. A possible channel that facilitates fast discharges may arise if the 
QCD top form mixes with another top form, either emergent or fundamental. As long as the charges of the membranes 
sourcing the new top form are comparable to their tensions, and the tensions are not too great, $\sim 3 keV$
the discharge processes could be fast enough, and remove the CP-violating phase rapidly. 

In this work we presented a
simple model showing how this can proceed, which could approximately 
restore CP by rapid cosmological membrane
production, which looks like very fast phase transitions around the chiral symmetry breaking scale.
The discharges should be fast since CP breaking occurs around the chiral symmetry breaking scale, which
is within few orders of magnitude from the BBN scale. To
play it safe the transitions 
should complete by BBN, maintaining the delicate balance between the various ingredients of the 
universe at that time. 

The price to pay is to have membranes with 
both charges and tensions being rather small. A small charge 
ensures that the discharge will be refined enough to yield the terminal $\theta_{\cancel{\tt CP}}$ below
$10^{-10}$, and a small membrane tension then makes the discharges fast, achieving  
$\theta_{\cancel{\tt CP}} \la 10^{-10}$ before BBN. Curiously, both criteria can be fitted naturally by 
picking a single scale for the membrane charge and tension, which is about $3 keV$. 
Obvious questions about these requirements is how well are such regimes under control, 
and how they might be realized. 

Regarding the smallness of charges and tensions, 
the former could be accomplished by e.g. kinetic mixing, similar to 
how the millicharge particles arise in field theory \cite{Holdom:1985ag}. 
Another way to do this is by UV completing the 
theory in some framework with compact large extra dimensions, which can make the final $4D$ value of charge 
of a membrane smaller than the fundamental scale since it involves a factor of compact internal space volume 
\cite{Feng:2000if}. It is also possible that there are multiple forms ${\cal H}_i$, with larger charges but with 
small differences between them.

The paper \cite{Feng:2000if} also points out a way how to make small tensions, by wrapping $p$-branes on 
shrinking cycles in internal tensions. However, recall that small tension is merely a simple way
to implement fast transitions. An alternative could be to have enhanced decay rates even without making
tensions small. This requires different decay channels for false vacua with $\theta_{\cancel{\tt CP}} > 10^{-10}$. 
Several different approaches were devised over time, including enhancements from thermal background effects
\cite{Garriga:2003gv,Brown:2015kgj}, favorable energetics 
of transition which enhance decay rates by favoring simultaneous
multi-membrane emissions \cite{Garriga:2003gv,Brown:2010bc,Brown:2010mg}, catalysis of fast transitions by
classical bubble collisions \cite{Easther:2009ft}, or rapid discharge cascades by unwiding of higher dimensional
objects that appear as membranes after compactification to $4D$ \cite{Kleban:2011cs}. Yet another possibility might also
be using heavier axions, which reside in their potential minima at chiral symmetry breaking, but where potentials
have a large number of vacua \cite{Bachlechner:2017zpb,Demirtas:2021gsq}. 
Some of these vacua might be `close' in field space, with enhanced tunneling rates, 
allowing for fast transitions. Those transitions could appear as rapid 
membrane nucleations if the axion multiverse were realized in ways 
which incorporate flux monodromies \cite{Kaloper:2023kua}. 

Given the existence of the flux discharge by membrane emissions, and the possibility to use it as an 
extension of the axion relaxation of $\theta_{\cancel{\tt CP}}$, it 
seems warranted to examine the flux discharge dynamics as a 
possible means to resolve the strong CP problem more closely. 

\vskip.3cm

{\bf Acknowledgments}: The author thanks A. Lawrence and 
A. Westphal for useful discussions, and G. Dvali for comments. The author is grateful to KITP UCSB for kind
hospitality in the course of this work. The research reported here 
was supported in part by the DOE Grant DE-SC0009999. This research
was also supported in part by grant NSF PHY-2309135 and the Gordon and Betty Moore
Foundation Grant No. 2919.02 to the Kavli Institute for Theoretical Physics (KITP). 



\begin{thebibliography}{99}

\bibitem{Weinberg:1978uk}
S.~Weinberg, ``Opening Address-Neutrinos '78,''
contribution to: Neutrino 78, 1-21, 
6th International Conference on Neutrino Physics, 
28 April 1978. Lafayette, IN, USA.

\bibitem{Aurilia:1980xj}
A.~Aurilia, H.~Nicolai and P.~K.~Townsend,
``Hidden Constants: The Theta Parameter of QCD and the Cosmological Constant of N=8 Supergravity,''
Nucl. Phys. B \textbf{176}, 509-522 (1980). 

\bibitem{Aurilia:1978qs}
A.~Aurilia, D.~Christodoulou and F.~Legovini,
``A Classical Interpretation of the Bag Model for Hadrons,''
Phys. Lett. B \textbf{73}, 429-432 (1978).

\bibitem{Duff:1980qv}
M.~J.~Duff and P.~van Nieuwenhuizen,
``Quantum Inequivalence of Different Field Representations,''
Phys. Lett. B \textbf{94}, 179-182 (1980). 

\bibitem{Coleman:1976uz}
S.~R.~Coleman,
``More About the Massive Schwinger Model,''
Annals Phys. \textbf{101}, 239 (1976). 

\bibitem{Rosenzweig:1979ay}
C.~Rosenzweig, J.~Schechter and C.~G.~Trahern,
``Is the Effective Lagrangian for QCD a Sigma Model?,''
Phys. Rev. D \textbf{21}, 3388 (1980). 

\bibitem{DiVecchia:1980yfw}
P.~Di Vecchia and G.~Veneziano,
``Chiral Dynamics in the Large n Limit,''
Nucl. Phys. B \textbf{171}, 253-272 (1980).

\bibitem{Nath:1979ik}
P.~Nath and R.~L.~Arnowitt,
``The U(1) Problem: Current Algebra and the Theta Vacuum,''
Phys. Rev. D \textbf{23}, 473 (1981).

\bibitem{Witten:1980sp}
E.~Witten,
``Large N Chiral Dynamics,''
Annals Phys. \textbf{128}, 363 (1980). 

\bibitem{Ohta:1981ai}
N.~Ohta,
``Vacuum Structure and Chiral Charge Quantization in the Large $N$ Limit,''
Prog. Theor. Phys. \textbf{66}, 1408 (1981)
[erratum: Prog. Theor. Phys. \textbf{67}, 993 (1982)].

\bibitem{Luscher:1978rn}
M.~L\"uscher, ``The Secret Long Range Force in Quantum Field Theories With Instantons,''
Phys. Lett. B \textbf{78}, 465-467 (1978).

\bibitem{DAdda:1978vbw}
A.~D'Adda, M.~Luscher and P.~Di Vecchia,
``A 1/n Expandable Series of Nonlinear Sigma Models with Instantons,''
Nucl. Phys. B \textbf{146}, 63-76 (1978).

\bibitem{Witten:1978bc}
E.~Witten,
``Instantons, the Quark Model, and the 1/n Expansion,''
Nucl. Phys. B \textbf{149}, 285-320 (1979). 

\bibitem{Dvali:2003br}
G.~Dvali and A.~Vilenkin,
``Cosmic attractors and gauge hierarchy,''
Phys. Rev. D \textbf{70}, 063501 (2004)
[arXiv:hep-th/0304043 [hep-th]].

\bibitem{Dvali:2004tma}
G.~Dvali,
``Large hierarchies from attractor vacua,''
Phys. Rev. D \textbf{74}, 025018 (2006)
[arXiv:hep-th/0410286 [hep-th]].

\bibitem{Dvali:2005an}
G.~Dvali, ``Three-form gauging of axion symmetries and gravity,''
[arXiv:hep-th/0507215 [hep-th]].

\bibitem{Dvali:2005zk}
G.~Dvali,
``A Vacuum accumulation solution to the strong CP problem,''
Phys. Rev. D \textbf{74}, 025019 (2006)
[arXiv:hep-th/0510053 [hep-th]].

\bibitem{Dvali:2007iv}
G.~Dvali and G.~R.~Farrar,
``Strong CP Problem with 10**32 Standard Model Copies,''
Phys. Rev. Lett. \textbf{101}, 011801 (2008)
[arXiv:0712.3170 [hep-th]].

\bibitem{Aurilia:1980jz}
A.~Aurilia, Y.~Takahashi and P.~K.~Townsend,
``The U(1) Problem and the Higgs Mechanism in Two-dimensions and Four-dimensions,''
Phys. Lett. B \textbf{95}, 265-268 (1980).

\bibitem{Jackiw:1976pf}
R.~Jackiw and C.~Rebbi,
``Vacuum Periodicity in a Yang-Mills Quantum Theory,''
Phys. Rev. Lett. \textbf{37}, 172-175 (1976).

\bibitem{Callan:1976je}
C.~G.~Callan, Jr., R.~F.~Dashen and D.~J.~Gross,
``The Structure of the Gauge Theory Vacuum,''
Phys. Lett. B \textbf{63}, 334-340 (1976).

\bibitem{Shifman:1998if}
M.~A.~Shifman,
``Domain walls and decay rate of the excited vacua in the large N Yang-Mills theory,''
Phys. Rev. D \textbf{59}, 021501 (1999)
[arXiv:hep-th/9809184 [hep-th]].

\bibitem{Forbes:2000et}
M.~M.~Forbes and A.~R.~Zhitnitsky, ``Domain walls in QCD,''
JHEP \textbf{10}, 013 (2001)
[arXiv:hep-ph/0008315 [hep-ph]].

\bibitem{Dvali:1998ms}
G.~R.~Dvali and Z.~Kakushadze,
``Large N domain walls as D-branes for N=1 QCD string,''
Nucl. Phys. B \textbf{537}, 297-316 (1999)
[arXiv:hep-th/9807140 [hep-th]].

\bibitem{Dubovsky:2011tu}
S.~Dubovsky, A.~Lawrence and M.~M.~Roberts,
``Axion monodromy in a model of holographic gluodynamics,''
JHEP \textbf{02}, 053 (2012)
[arXiv:1105.3740 [hep-th]].

\bibitem{Kaloper:2025wgn}
N.~Kaloper, ``An Alternative to Axion,''
[arXiv:2504.21078 [hep-ph]].

\bibitem{Kosowsky:1992rz} 
A.~Kosowsky, M.~S.~Turner and R.~Watkins,
``Gravitational waves from first order cosmological phase transitions,''
Phys. Rev. Lett. \textbf{69}, 2026-2029 (1992). 

\bibitem{Kamionkowski:1993fg}
M.~Kamionkowski, A.~Kosowsky and M.~S.~Turner,
``Gravitational radiation from first order phase transitions,''
Phys. Rev. D \textbf{49}, 2837-2851 (1994)
[arXiv:astro-ph/9310044 [astro-ph]].

\bibitem{Veneziano:1979ec}
G.~Veneziano,
``U(1) Without Instantons,''
Nucl. Phys. B \textbf{159}, 213-224 (1979).

\bibitem{Kogut:1974kt}
J.~B.~Kogut and L.~Susskind,
``How to Solve the eta --\ensuremath{>} 3 pi Problem by Seizing the Vacuum,''
Phys. Rev. D \textbf{11}, 3594 (1975). 

\bibitem{Vafa:1984xg}
C.~Vafa and E.~Witten,
``Parity Conservation in QCD,''
Phys. Rev. Lett. \textbf{53}, 535 (1984). %

\bibitem{Polyakov:1975yp}
A.~M.~Polyakov and A.~A.~Belavin, ``Metastable States of Two-Dimensional Isotropic Ferromagnets,''
JETP Lett. \textbf{22}, 245-248 (1975). 

\bibitem{DAdda:1978dle}
A.~D'Adda, P.~Di Vecchia and M.~L\"uscher, 
``Confinement and Chiral Symmetry Breaking in $CP^{n-1}$ Models with Quarks,''
Nucl. Phys. B \textbf{152}, 125-144 (1979). 

\bibitem{Kaloper:2016fbr}
N.~Kaloper and A.~Lawrence,
``London equation for monodromy inflation,''
Phys. Rev. D \textbf{95}, no.6, 063526 (2017)
[arXiv:1607.06105 [hep-th]].

\bibitem{Gabadadze:1997kj}
G.~Gabadadze,
``Modeling the glueball spectrum by a closed bosonic membrane,''
Phys. Rev. D \textbf{58}, 094015 (1998)
[arXiv:hep-ph/9710402 [hep-ph]].

\bibitem{Gabadadze:2002ff}
G.~Gabadadze and M.~Shifman,
``QCD vacuum and axions: What's happening?,''
Int. J. Mod. Phys. A \textbf{17}, 3689-3728 (2002)
[arXiv:hep-ph/0206123 [hep-ph]].

\bibitem{Aurilia:1978dw}
A.~Aurilia, ``The Problem of Confinement: From Two-dimensions to Four-dimensions,''
Phys. Lett. B \textbf{81}, 203-206 (1979). 

\bibitem{Kaloper:2022oqv}
N.~Kaloper,
``Hidden variables of gravity and geometry and the cosmological constant problem,''
Phys. Rev. D \textbf{106}, no.6, 065009 (2022)
[arXiv:2202.06977 [hep-th]].

\bibitem{Kaloper:2022utc}
N.~Kaloper,
``Pancosmic Relativity and Nature's Hierarchies,''
Phys. Rev. D \textbf{106}, no.4, 044023 (2022)
[arXiv:2202.08860 [hep-th]].

\bibitem{Peccei:1977hh}
R.~D.~Peccei and H.~R.~Quinn,
``CP Conservation in the Presence of Instantons,''
Phys. Rev. Lett. \textbf{38}, 1440-1443 (1977).

\bibitem{Weinberg:1977ma}
S.~Weinberg,
``A New Light Boson?,''
Phys. Rev. Lett. \textbf{40}, 223-226 (1978).

\bibitem{Wilczek:1977pj}
F.~Wilczek,
``Problem of Strong  $P$  and  $T$  Invariance in the Presence of Instantons,''
Phys. Rev. Lett. \textbf{40}, 279-282 (1978).

\bibitem{Kim:1979if}
J.~E.~Kim,
``Weak Interaction Singlet and Strong CP Invariance,''
Phys. Rev. Lett. \textbf{43}, 103 (1979). 

\bibitem{Shifman:1979if}
M.~A.~Shifman, A.~I.~Vainshtein and V.~I.~Zakharov,
``Can Confinement Ensure Natural CP Invariance of Strong Interactions?,''
Nucl. Phys. B \textbf{166}, 493-506 (1980). 

\bibitem{Zhitnitsky:1980tq}
A.~R.~Zhitnitsky,
``On Possible Suppression of the Axion Hadron Interactions. (In Russian),''
Sov. J. Nucl. Phys. \textbf{31}, 260 (1980). 

\bibitem{Dine:1981rt}
M.~Dine, W.~Fischler and M.~Srednicki,
``A Simple Solution to the Strong CP Problem with a Harmless Axion,''
Phys. Lett. B \textbf{104}, 199-202 (1981). 

\bibitem{Kaloper:2008qs}
N.~Kaloper and L.~Sorbo,
``Where in the String Landscape is Quintessence,''
Phys. Rev. D \textbf{79}, 043528 (2009)
[arXiv:0810.5346 [hep-th]].

\bibitem{Kaloper:2008fb}
N.~Kaloper and L.~Sorbo,
``A Natural Framework for Chaotic Inflation,''
Phys. Rev. Lett. \textbf{102}, 121301 (2009)
[arXiv:0811.1989 [hep-th]].

\bibitem{Kaloper:2011jz}
N.~Kaloper, A.~Lawrence and L.~Sorbo,
``An Ignoble Approach to Large Field Inflation,''
JCAP \textbf{03}, 023 (2011)
[arXiv:1101.0026 [hep-th]].

\bibitem{Julia:1979ur}
B.~Julia and G.~Toulouse,
``The Many Defect Problem: Gauge Like Variables for Ordered Media Containing Defects,''
J. Phys. Lett. \textbf{40}, 396 (1979). 

\bibitem{Quevedo:1996uu}
F.~Quevedo and C.~A.~Trugenberger,
``Phases of antisymmetric tensor field theories,''
Nucl. Phys. B \textbf{501}, 143-172 (1997)
[arXiv:hep-th/9604196 [hep-th]].

\bibitem{Gnadig:1976pn}
P.~Gnadig, P.~Hasenfratz, J.~Kuti and A.~S.~Szalay,
``The Quark Bag Model with Surface Tension,''
Phys. Lett. B \textbf{64}, 62-66 (1976).

\bibitem{Brown:1987dd}
J.~D.~Brown and C.~Teitelboim,
``Dynamical Neutralization of the Cosmological Constant,''
Phys. Lett. B \textbf{195}, 177-182 (1987). 

\bibitem{Brown:1988kg}
J.~D.~Brown and C.~Teitelboim,
``Neutralization of the Cosmological Constant by Membrane Creation,''
Nucl. Phys. B \textbf{297}, 787-836 (1988). 

\bibitem{Schwinger:1951nm}
J.~S.~Schwinger,
``On gauge invariance and vacuum polarization,''
Phys. Rev. \textbf{82}, 664-679 (1951).


\bibitem{Coleman:1977py}
S.~R.~Coleman,
``The Fate of the False Vacuum. 1. Semiclassical Theory,''
Phys. Rev. D \textbf{15}, 2929-2936 (1977)
[erratum: Phys. Rev. D \textbf{16}, 1248 (1977)].

\bibitem{Callan:1977pt}
C.~G.~Callan, Jr. and S.~R.~Coleman,
``The Fate of the False Vacuum. 2. First Quantum Corrections,''
Phys. Rev. D \textbf{16}, 1762-1768 (1977). 

\bibitem{Garriga:1993fh}
J.~Garriga,
``Nucleation rates in flat and curved space,''
Phys. Rev. D \textbf{49}, 6327-6342 (1994)
[arXiv:hep-ph/9308280 [hep-ph]].

\bibitem{Guth:1982pn}
A.~H.~Guth and E.~J.~Weinberg,
``Could the Universe Have Recovered from a Slow First Order Phase Transition?,''
Nucl. Phys. B \textbf{212}, 321-364 (1983).

\bibitem{Turner:1992tz}
M.~S.~Turner, E.~J.~Weinberg and L.~M.~Widrow,
``Bubble nucleation in first order inflation and other cosmological phase transitions,''
Phys. Rev. D \textbf{46}, 2384-2403 (1992).

\bibitem{Freese:2004vs}
K.~Freese and D.~Spolyar,
``Chain inflation: 'Bubble bubble toil and trouble',''
JCAP \textbf{07}, 007 (2005)
[arXiv:hep-ph/0412145 [hep-ph]].

\bibitem{Zeldovich:1974uw}
Y.~B.~Zeldovich, I.~Y.~Kobzarev and L.~B.~Okun,
``Cosmological Consequences of the Spontaneous Breakdown of Discrete Symmetry,''
Zh. Eksp. Teor. Fiz. \textbf{67}, 3-11 (1974). 

\bibitem{Kim:1986ax}
J.~E.~Kim,
``Light Pseudoscalars, Particle Physics and Cosmology,''
Phys. Rept. \textbf{150}, 1-177 (1987). 

\bibitem{Berezin:1987bc}
V.~A.~Berezin, V.~A.~Kuzmin and I.~I.~Tkachev,
``Dynamics of Bubbles in General Relativity,''
Phys. Rev. D \textbf{36}, 2919 (1987). 

\bibitem{Banks:1991mb}
T.~Banks, M.~Dine and N.~Seiberg,
``Irrational axions as a solution of the strong CP problem in an eternal universe,''
Phys. Lett. B \textbf{273}, 105-110 (1991)
[arXiv:hep-th/9109040 [hep-th]].

\bibitem{Anastassopoulos:2015ura}
V.~Anastassopoulos, S.~Andrianov, R.~Baartman, M.~Bai, S.~Baessler, J.~Benante, M.~Berz, M.~Blaskiewicz, T.~Bowcock and K.~Brown, \textit{et al.}
``A Storage Ring Experiment to Detect a Proton Electric Dipole Moment,''
Rev. Sci. Instrum. \textbf{87}, no.11, 115116 (2016)
[arXiv:1502.04317 [physics.acc-ph]].

\bibitem{Zhevlakov:2020bvr}
A.~S.~Zhevlakov and V.~E.~Lyubovitskij,
``Deuteron EDM induced by $CP$ violating couplings of pseudoscalar mesons,''
Phys. Rev. D \textbf{101}, no.11, 115041 (2020)
[arXiv:2003.12217 [hep-ph]].

\bibitem{Kharzeev:1998kz}
D.~Kharzeev, R.~D.~Pisarski and M.~H.~G.~Tytgat,
``Possibility of spontaneous parity violation in hot QCD,''
Phys. Rev. Lett. \textbf{81}, 512-515 (1998)
[arXiv:hep-ph/9804221 [hep-ph]].

\bibitem{Halperin:1998gx}
I.~E.~Halperin and A.~Zhitnitsky,
``Axion potential, topological defects and CP odd bubbles in QCD,''
Phys. Lett. B \textbf{440}, 77-88 (1998)
[arXiv:hep-ph/9807335 [hep-ph]].

\bibitem{Kharzeev:1998kya}
D.~Kharzeev, R.~D.~Pisarski and M.~H.~G.~Tytgat, ``Parity odd bubbles in hot QCD,''
[arXiv:hep-ph/9808366 [hep-ph]].

\bibitem{Buckley:1999mv}
K.~Buckley, T.~Fugleberg and A.~Zhitnitsky,
``Can theta vacua be created in heavy ion collisions?,''
Phys. Rev. Lett. \textbf{84}, 4814-4817 (2000). 
[arXiv:hep-ph/9910229 [hep-ph]].

\bibitem{Kharzeev:2007tn}
D.~Kharzeev and A.~Zhitnitsky,
``Charge separation induced by P-odd bubbles in QCD matter,''
Nucl. Phys. A \textbf{797}, 67-79 (2007)
[arXiv:0706.1026 [hep-ph]].

\bibitem{Holdom:1985ag}
B.~Holdom,
``Two U(1)'s and Epsilon Charge Shifts,''
Phys. Lett. B \textbf{166}, 196-198 (1986). 

\bibitem{Feng:2000if}
J.~L.~Feng, J.~March-Russell, S.~Sethi and F.~Wilczek,
``Saltatory relaxation of the cosmological constant,''
Nucl. Phys. B \textbf{602}, 307-328 (2001) 
[arXiv:hep-th/0005276 [hep-th]].

\bibitem{Garriga:2003gv}
J.~Garriga and A.~Megevand,
``Coincident brane nucleation and the neutralization of Lambda,''
Phys. Rev. D \textbf{69}, 083510 (2004)
[arXiv:hep-th/0310211 [hep-th]].

\bibitem{Brown:2015kgj}
A.~R.~Brown,
``Schwinger pair production at nonzero temperatures or in compact directions,''
Phys. Rev. D \textbf{98}, no.3, 036008 (2018)
[arXiv:1512.05716 [hep-th]].

\bibitem{Brown:2010bc}
A.~R.~Brown and A.~Dahlen,
``Small Steps and Giant Leaps in the Landscape,''
Phys. Rev. D \textbf{82}, 083519 (2010)
[arXiv:1004.3994 [hep-th]].

\bibitem{Brown:2010mg}
A.~R.~Brown and A.~Dahlen,
``Giant Leaps and Minimal Branes in Multi-Dimensional Flux Landscapes,''
Phys. Rev. D \textbf{84}, 023513 (2011)
[arXiv:1010.5241 [hep-th]].

\bibitem{Easther:2009ft}
R.~Easther, J.~T.~Giblin, Jr, L.~Hui and E.~A.~Lim,
``A New Mechanism for Bubble Nucleation: Classical Transitions,''
Phys. Rev. D \textbf{80}, 123519 (2009)
[arXiv:0907.3234 [hep-th]].

\bibitem{Kleban:2011cs}
M.~Kleban, K.~Krishnaiyengar and M.~Porrati,
``Flux Discharge Cascades in Various Dimensions,''
JHEP \textbf{11}, 096 (2011)
[arXiv:1108.6102 [hep-th]].

\bibitem{Bachlechner:2017zpb}
T.~C.~Bachlechner, K.~Eckerle, O.~Janssen and M.~Kleban,
``Multiple-axion framework,''
Phys. Rev. D \textbf{98}, no.6, 061301 (2018)
[arXiv:1703.00453 [hep-th]].

\bibitem{Demirtas:2021gsq}
M.~Demirtas, N.~Gendler, C.~Long, L.~McAllister and J.~Moritz,
``PQ axiverse,''
JHEP \textbf{06}, 092 (2023)
[arXiv:2112.04503 [hep-th]].

\bibitem{Kaloper:2023kua}
N.~Kaloper,
``Axion flux monodromy discharges relax the cosmological constant,''
JCAP \textbf{11}, 032 (2023)
[arXiv:2307.10365 [hep-th]].

\end{thebibliography}
\end{document}